\documentclass{ptephy_v1}

\usepackage{graphicx}
\usepackage{bm}
\usepackage{amsmath}
\usepackage{amssymb}
\usepackage{mathtools}
\usepackage{cancel}
\usepackage{ascmac}
\usepackage{fancybox}
\usepackage{color}
\usepackage{latexsym}

\def\Nu#1#2#3{{}^{#2}_{#3}\mathrm{#1}}

\newcommand{\threeDGraphs}[3]{
\vspace{-10mm}
\hspace{-9mm}
\begin{tabular}{ccc}
\includegraphics[width=0.33\columnwidth] {#1}
&
\hspace{-2mm}
\includegraphics[width=0.33\columnwidth] {#2}
&
\hspace{-2mm}
\includegraphics[width=0.33\columnwidth] {#3}
\end{tabular}
}

\def\beq{\begin{equation}}
\def\eeq{\end{equation}}
\def\beqn{\begin{eqnarray*}}
\def\eeqn{\end{eqnarray*}}
\def\beqnn{\begin{eqnarray}}
\def\eeqnn{\end{eqnarray}}

\begin{document}
\title{General theory of constructing potential with bound states in the continuum}


\author{Mao Kurino}
\author{Kazuo Takayanagi}
\affil{Department of Physics, Sophia University,
7-1 Kioi-cho, Chiyoda-ku, Tokyo 102, Japan \email{m-kurino-ex5@eagle.sophia.ac.jp, k-takaya@sophia.ac.jp}}





\begin{abstract}
We present a general theory of  potentials that support
bound states at positive energies (bound states in the continuum).
On the theoretical side, we prove that,  for systems described by 
 nonlocal potentials of the form $V(r,r')$,  bound states at positive energies 
are as common as those at negative energies.  
At the same time, we show that a local potential of the form $V(r)$ rarely supports
a positive energy bound state.
On the practical side, we show how to construct a (naturally nonlocal) potential 
 which supports an arbitrary normalizable state at an arbitrary positive energy.
We demonstrate our theory with numerical examples both in momentum and coordinate spaces
with emphasis on the important role played by nonlocal potentials.
Finally, we discuss how to observe 
 bound states at positive energies, and  where to search for 
nonlocal potentials which may support them. 
\end{abstract}

\subjectindex{A64, D00, D06, D13, D14}

\maketitle

\section{Introduction}
\label{sec:intro}

Since the pioneering work of von Neumann and Wigner \cite{vonNeumann29},
a lot of attempts have been made both theoretically and experimentally 
in search of  bound states at positive energies, 
or equivalently   bound states  in the continuum (BIC's) in quantum systems.
Theoretically,  the BIC has been described as 
 a resonance with vanishing width which  occurs
 only  accidentally 
\cite{Stillinger_PRA_75, Pursey_95, Weber_98}.
On the experimental side, however, 
no clear BIC has been observed experimentally in quantum systems, 
while  many BIC's have been found in a broad range of classical 
and semi-classical systems
\cite{Plotnik_11, Cerjan_19}.

In this situation, we present a general theory of the potential which supports a BIC
via a simple one-body Schr\"odinger equation.\footnote{
In this work, we are concerned with localized potentials with which we are familiar 
in nuclear and atomic physics.
Therefore, the Anderson localization in solid state physics, for example, 
is not relevant in the present work  though 
it gives a localized state at a positive energy.
This is because it requires a random potential 
which is not localized but exists all over the system.
}
We express a Hermitian potential as $V=V_{\rm S}+V_{\rm B}$,
which we call  SB-decomposition  \cite{Takayanagi_PTEP_23}, where
$V_{\rm S}$ and $V_{\rm B}$ operate
 on the spaces of scattering and bound states, respectively.
Then, by examining  transformations of $V_{\rm S}$ and $V_{\rm B}$ separately, 
we show how to design a Hermitian potential 
which describes an arbitrary normalizable state
as a BIC at an arbitrary positive energy.
 Conversely, any Hermitian potential supporting a BIC can be described 
by the present theory. 
In other words, we explicitly present a general solution to
the inverse problem for an arbitrary BIC.

We develop the theory in momentum space, i.e., we construct a 
BIC-supporting potential in the form of $\langle k| V| k' \rangle$ where
$k$ and $k'$ are the relative momenta for a fixed angular momentum.
By taking its Fourier transform, we obtain a BIC-supporting potential
$V(r,r') = \langle r| V| r' \rangle$ in coordinate space 
which generally depends on $r$ and $r'$ separately, 
and is therefore nonlocal.
In the present work,  it is important to distinguish 
between a familiar ``local''  potential $V(r)$  
 and a general ``nonlocal'' potential $V(r,r')$.\footnote{
A Hermitian potential is generally a nonlocal potential  $V(r,r') = \langle r| V| r' \rangle$
in the coordinate representation.
A nonlocal potential $V(r,r')$ enters Schr\"odinger equation for $\phi(r)$ as
$\displaystyle{ \int_0^{\infty} \! r'^2{\rm d}r'  \, V(r,r') \phi (r')}$
as in Eq.(\ref{eqn:Sch_general_coordinate}),
while a local potential $V(r)$ shows up in the form 
of $V(r)\phi(r)$ as in Eq.(\ref{eqn:Sch_general_coordinate_local}).
Note that a local potential  $V(r)$ can be expressed as a nonlocal potential
as $V(r,r') = \dfrac{\delta(r'-r)}{r'r} V(r)$, 
meaning that a local potential 
is a special case of a nonlocal potential.
As a simple example of the nonlocal potential, one may think of 
the Fock potential in the Hartree-Fock theory.   
} 
 
The plan of this paper is as follows.
In Sect.\,\ref{sec:general_design_V}, we briefly explain the general theory to design 
 a Hermitian potential $V$
using the SB-decomposition.  
Then, in  Sect.\,\ref{sec:V_with_BIC}, we present two designs of  
BIC-supporting potentials.  
One design requires specification of both $V_{\rm S}$ and $V_{\rm B}$,
and  is most general but demands sizable calculations.
The other uses transformation of $V_{\rm B}$ only, 
and  is quite simple and numerically feasible without complicated calculations. 
In Sects.\,\ref{sec:numerical_example_simple} and \ref{sec:numerical_example_general},
 we show numerical results 
which demonstrate our designs of BIC-supporting potentials 
in both momentum and coordinate spaces.   Here, we shall realize
the important role played by the nonlocality of potential to support a BIC.   
In Sect.\,\ref{sec:discussion_local},
we discuss the BIC from the viewpoint of familiar local potentials,
and prove that, 
for a local potential $V(r)$ to describe a  bound state wave function $\phi(r)$ as a BIC, 
both $V(r)$ and $\phi(r)$ must be oscillating functions 
which decay very slowly as $r\to \infty$.
Therefore, a local potential  can rarely support a BIC.
In Sects.\,\ref{sec:discussion} and \ref{sec:local_vs_nonlocal}, 
we discuss the BIC using  nonlocal potentials 
on the basis of both the Schr\"odinger  equation for a bound state and 
the  structure of the group of  all Hermitian potentials with a bound state.
 Surprisingly enough, 
the present theory clarifies that,
in systems described by Hermitian nonlocal potentials in general,
  positive energy bound states (BIC's)
are as common a phenomenon as negative energy bound states. 
To find a BIC, therefore, 
it is crucial to examine systems with nonlocal potentials.
In Sect.\,\ref{sec:observing_BIC}, 
we show how we can observe a BIC, if exists, 
on the basis of Levinson's theorem and $\langle k| V_{\rm B}| k' \rangle$ 
 both theoretically and experimentally.
Then,  in Sect.\,\ref{sec:quest_BIC}, we discuss how and where to
search for a BIC in practice.   Here, facing the problem of 
nonlocal potentials which are not commonplace in many fields of physics, 
we present possible scenarios to look for a BIC in real physical systems.   
Finally  in Sect.\,\ref{sec:summary}, we summarize the present work.

\section{General design of potential}
\label{sec:general_design_V}
We consider a spinless particle of mass $m$ described by the 
 Hamiltonian:  
\begin{equation}
H = H_0  + V ,
\label{eqn:def_H}
\end{equation} 
where $H_0$ is the kinetic energy, and $V$ is the potential energy
of the particle with a fixed force center.

In the following, we briefly explain our method to construct a potential 
$V$ in three steps (i), (ii), and (iii) in a way which clarifies 
 its relation to the bound and scattering states.
Our method is based on the SB-decomposition
 of a Hermitian potential $V$, which derives from the generalized
optical theorem for the half-on-shell T-matrix \cite{Takayanagi_PTEP_23}.
In what follows,   we assume a rotationally invariant system for simplicity, 
and work in the space of a fixed partial wave.

\begin{enumerate}
\item[(i)] -- ${\mathcal B}$

\noindent
First, in the whole Hilbert space  ${\mathcal H}$,
we specify an $N$-dimensional subspace ${\mathcal B}$  which is to be
spanned by $N$ bound states we are going to fix below.
Naturally, its orthogonal complement ${\mathcal S}$ is
to be spanned by as-yet-unknown scattering states.   
With knowledge of the subspace ${\mathcal B}$, we can define 
 the projection operators $P_{\rm B} $ onto ${\mathcal B}$,\footnote{
Once the $N$-dimensional subspace ${\mathcal B}$ is fixed, 
any orthonormal set $\{ | \chi_i \rangle , i = 1,2,\cdots,N \}$ 
satisfying ${\mathcal B}= {\rm Span} \,\{ | \chi_i \rangle \}$
expresses  the projection operator $P_{\rm B}$ as
 $P_{\rm B} = \sum_{i}   | \chi_i \rangle  \langle  \chi_i |$.
}    
and $P_{\rm S}=1-P_{\rm B} $
 onto ${\mathcal S}$.
Then, we decompose the Hermitian potential $V$ we are going to construct as 
\begin{equation}
V = V_{\rm S} + V_{\rm B},
\label{eqn:V_eq_VPB_plus_VPS}
\end{equation}
where we have used abbreviations
$VP_{\rm S} \rightarrow V_{\rm S}$ and $ VP_{\rm B} \rightarrow V_{\rm B}$.
In the following, Eq.(\ref{eqn:V_eq_VPB_plus_VPS}) is referred to as
the SB-decomposition of $V$ for  obvious reasons.  
\vspace{2mm}

\item[(ii)] -- $V_{\rm S}$

\noindent
Second, for the fixed subspace ${\mathcal B}$, 
the generalized optical theorem states the following 
\cite{Takayanagi_PTEP_23,Takayanagi_JMP_15, Takayanagi_PRA_08}; 
 a function $\langle k' | T  | k \rangle $
of two variables $k'$ and $k$ is physically acceptable
as a half-on-shell T-matrix,\footnote{
The half-on-shell T-matrix $\langle k' |T| k \rangle$ is defined via the Lippmann-Schwinger equation \cite{Newton82,Taylor83}, 
${\displaystyle
\langle k' |T| k \rangle = \langle k' |V| k \rangle + 
  \int \!\! \dfrac{p^2 {\rm d}p}{(2\pi)^3}
~\langle k' | V  | p  \rangle \dfrac{1}{k^2-p^2 + i\eta}   \langle p | T  | k \rangle 
}$. 
Note that we adopt the usual unit system where $\hbar = 2m =1$.
}
 i.e., it has an underlying Hermitian potential $V$,
 if and only if it satisfies following conditions  ({\bf A}) and ({\bf B}).
\begin{align}
T\Omega_{+}^{\dagger} - H_0 P_{\rm B} 
&= \Omega_{+} T^{\dagger} - P_{\rm B} H_0 ,
\tag{{\bf A}}
\\
\Omega_{+}^{\dagger} T &=  T^{\dagger} \Omega_{+},
\tag{{\bf B}}
\end{align}
where $\Omega_{+}$ is the M{\o}ller wave operator  that transforms 
a plane wave state $| k \rangle$ into its corresponding scattering state 
$| k \rangle\!_+ = \Omega_{+} | k \rangle$
with the outgoing wave boundary condition \cite{Newton82, Taylor83}.\footnote{
$\Omega_{+}$ is defined in terms of $T$ as 
$\langle p |\Omega_{+}| k \rangle
=  \dfrac{(2\pi)^3}{pk} \delta(p-k)+ \dfrac{1}{k^2-p^2 + i\eta}   \langle p | T  | k \rangle$. 
Therefore,  conditions  ({\bf A}) and ({\bf B})
compose a coupled set of equations for $ \langle p | T  | k \rangle$,
a function  of two variables $p$ and $k$,  
to be compatible with the given $P_{\rm B}$.
}
There are many $T$'s
that satisfy  conditions ({\bf A}) and  ({\bf B})
 for the fixed subspace ${\mathcal B}$.
Therefore, we arbitrarily single out a $T$ from them, 
which in turn defines its corresponding $V_{\rm S}$ as
\begin{equation}
\langle k' |V_{\rm S} | k  \rangle
=  \langle k' | T  | k  \rangle + 
\int \!\! \frac{p^2 {\rm d}p}{(2\pi)^3}
~\langle k' | T  | p  \rangle
     \frac{1}{p^2-k^2- i\eta} \langle p | T^{\dagger}  | k  \rangle.
\label{eqn:def_VPS}
\end{equation}
Then, it is easy to show that $V_{\rm S}$ of Eq.(\ref{eqn:def_VPS})
describes our scattering states  $\{ | k \rangle\!_+ = \Omega_{+} | k \rangle\}$ via 
the following Schr\"odinger  equation;
\begin{equation}
(H_0 +V) | k \rangle\!_{+} 
= (H_0+V_{\rm S}) | k \rangle\!_{+} 
= k^2 | k \rangle\!_{+}  .
\label{eqn:Sch_scat}
\end{equation}

\pagebreak
\item[(iii)]  -- $V_{\rm B}$

\noindent
Third, we arbitrarily specify an orthonormal set of bound states
$ \{ | \phi_i \rangle , i = 1,2,\cdots,N \}$
which span the subspace ${\mathcal B}$, i.e.,
  ${\mathcal B}= {\rm Span} \,\{ | \phi_i \rangle \}$.
Furthermore, we arbitrarily allot eigenenergy $E_i$ to $ | \phi_i \rangle$.
Then, it is straightforward to show that\footnote{
As in the standard notation,
$\langle k | \phi \rangle = \phi(k)$ 
is the wave function in momentum space.
}  
\begin{equation}
\langle k' |V_{\rm B} | k  \rangle
= \sum_{i} (E_i - {k'}^2) 
     \langle k'|\phi_i \rangle  \langle \phi_i  | k  \rangle,
\label{eqn:def_VPB2}
\end{equation}
describes our bound states $ \{ | \phi_i \rangle \}$ via
the  Schr\"odinger  equation, 
\begin{equation}
(H_0 +V) | \phi_i \rangle 
= (H_0+V_{\rm B})
 | \phi_i \rangle = E_{i}| \phi_i \rangle .
\label{eqn:Sch_bound}
\end{equation}

\end{enumerate}
\noindent
Here, conditions   ({\bf A}) and ({\bf B}) of the generalized optical theorem guarantee that,
 for $V_{\rm S}$ of Eq.(\ref{eqn:def_VPS}) and $V_{\rm B}$ of Eq.(\ref{eqn:def_VPB2}),
$V$ of Eq.(\ref{eqn:V_eq_VPB_plus_VPS}) is Hermitian \cite{Takayanagi_PTEP_23}.
Conversely,  any Hermitian potential $V$ can be uniquely SB-decomposed 
 into the form of Eq.(\ref{eqn:V_eq_VPB_plus_VPS}).
To summarize, above steps (i), (ii), and (iii) compose a general method
to design a Hermitian potential $V$ with keeping its eigenstates
 $\{ | k \rangle\!_+ ,\, k^2 \}$  and $ \{ | \phi_i \rangle,\, E_i \}$ under control.\footnote{
For convenience, we present the eigenstate $| \phi_i \rangle$ and
its eigenenergy $E_i$ in a single pair of braces.
}

We stress that, once the subspace ${\mathcal B}$ is fixed in step (i),
the SB-decomposition determines 
$V_{\rm S}$ and $V_{\rm B}$  quite independently
via steps (ii)  and (iii), respectively. 
This is because $V_{\rm S}$ and $V_{\rm B}$ depend on the bound states 
in different ways;  $V_{\rm B}$ of Eq.(\ref{eqn:def_VPB2}) 
demands all bound states 
$ \{ | \phi_i \rangle \}$
and their energies $ \{  E_i \}$ for its description.
On the other hand, $V_{\rm S}$ of Eq.(\ref{eqn:def_VPS}), 
which expresses  $V_{\rm S}$ in terms of $T$, 
requires the subspace ${\mathcal B}$ only, i.e.,  
it does not demand further information on bound states for its description.\footnote{ 
This is because bound states constrain $T$ (and therefore $V_{\rm S}$)
through conditions  ({\bf A}) and ({\bf B}) which 
 utilize the information of 
bound states via $P_{\rm B}$ only.
}
As we shall see, the above feature of the SB-decomposition suits 
the study of BIC.

\section{Potential with BIC}
\label{sec:V_with_BIC}
Using the three steps  in Sect.\,\ref{sec:general_design_V}, 
we explain how to construct a  BIC-supporting  potential.
After giving an explanation of  the most general method 
in Sect.\,\ref{sec:general_form_V_supporting_BIC},
we present a simple and practical method in Sect.\,\ref{sec:simple_form_V_supporting_BIC}.

\subsection{General method -- specification of  $V_{\rm S}$ and $V_{\rm B}$}
\label{sec:general_form_V_supporting_BIC}

Let us note that the SB-decomposition  in Sect.\,\ref{sec:general_design_V} 
does not require $E_i$ be negative in step (iii).
This implies that we already have a general design of the potential supporting a BIC;
by giving a positive value to $E_i$ in step (iii),
 our potential $V$ of Eq.(\ref{eqn:V_eq_VPB_plus_VPS}) describes  
 the bound state   $| \phi_i \rangle$ as a BIC.
Therefore,  via steps (i), (ii), and (iii) in Sect.\,\ref{sec:general_design_V}, 
we can construct a potential that supports an arbitrary
normalizable state $|\phi \rangle $ as a BIC 
at an arbitrary positive energy.
Conversely, because of the uniqueness of 
the SB-decomposition, 
any  Hermitian potential supporting a BIC 
 can surely be expressed uniquely  by 
Eqs.(\ref{eqn:V_eq_VPB_plus_VPS}), (\ref{eqn:def_VPS}), 
and (\ref{eqn:def_VPB2}).

\subsection{Simple method -- transformation of $V_{\rm B}$}
\label{sec:simple_form_V_supporting_BIC}

In the general method in Sect.\,\ref{sec:general_form_V_supporting_BIC}, 
we can arbitrarily specify the wave function and the eigenenergy 
of the BIC in steps (i) and (iii).   
 In trade for these  degrees of freedom, however, 
we have to look for a solution $\langle k' | T  | k  \rangle$ 
to the set of Eqs.({\bf A}) and ({\bf B}) in step (ii), 
which is sometimes inconvenient.    
In this section, therefore, 
we show a simple version of the general method,  
which  does not require  
an explicit solution to  Eqs.({\bf A}) and ({\bf B}).
We shall find that the present method is 
quite useful not only practically but also theoretically.\footnote{
We shall  make use of  the present simple method 
not only in actual calculations in Sect.\,\ref{sec:numerical_example_simple}
but also in theoretical arguments in Sect.\,\ref{sec:generality_BIC}.
}

\subsubsection{Derivation}
\label{sec:simple_form_V_supporting_BIC_1}

Here, we start with an arbitrary Hermitian potential $V_0$ 
which supports, for simplicity,  a single bound state
$| \phi \rangle$ at a negative energy  $E=-\gamma_0^2$;
\begin{equation}
(H_0 +V_0) | \phi \rangle 
 = -\gamma_0^2 | \phi \rangle .
\label{eqn:Sch_bound_simple_method}
\end{equation}
In the following, we derive a potential $V$ with a BIC by establishing 
 a transformation $V_0 \to V$
using the  three steps in  Sect.\,\ref{sec:general_design_V},
which we shall find quite simple and feasible.

\begin{enumerate}
\item[(i)]
We adopt $| \phi \rangle$ of Eq.(\ref{eqn:Sch_bound_simple_method})
as  the wave function of the BIC; 
instead of giving arbitrarily a wave function of the BIC,
we give arbitrarily a Hermitian potential $V_0$ and then adopt 
its bound state  $| \phi \rangle$ as the BIC
of the as-yet-unknown potential $V$  
which describes $| \phi \rangle$ via 
\begin{equation}
(H_0 +V) | \phi \rangle 
 = E | \phi \rangle , 
\label{eqn:Sch_BIC_formal}
\end{equation}
with a positive $E$.
The present choice of the BIC means that  
${\mathcal B}= {\rm Span} \,\{ | \phi \rangle \}$ of $V$ 
is the same as for $V_0$.
Therefore, their SB-decompositions, 
 $V=V_{\rm S} + V_{\rm B}$  and $V_0=V_{0,{\rm S}} + V_{0,{\rm B}}$,
are defined in terms of the same projection operators
 $P_{\rm B}$ and $P_{\rm S}$. 

\item[(ii)]
Let $T$ and $T_0$ be the half-on-shell T-matrices of $V$ and $V_0$, respectively.
Because ${\mathcal B}$ of $V$ 
is the same as for $V_0$, conditions  ({\bf A}) and ({\bf B}) for $T$ are the same 
as for $T_0$.
We can choose, therefore,  a trivial  solution of  Eqs.\,({\bf A}) and ({\bf B})
 without solving them for $T$, i.e., $T= T_0$, 
yielding $V_{\rm S}=V_{0,{\rm S}}$ via Eq.(\ref{eqn:def_VPS}).
Then, because the  SB-decomposition of $V_0$ gives $V_{0,{\rm B}}$ as
\begin{equation}
\langle k' |V_{0,{\rm B}} | k  \rangle
=  (-\gamma_0^2 - {k'}^2) 
     \langle k'|\phi \rangle  \langle \phi | k  \rangle,
\label{eqn:def_V0PB_single_BIC}
\end{equation}
 it immediately yields $V_{\rm S}=V_{0,{\rm S}} = V_0- V_{0,{\rm B}}$ as
\begin{equation}
\langle k' |V_{\rm S} | k  \rangle
=  \langle k' |V_0 | k  \rangle
+(\gamma_0^2 + {k'}^2) 
     \langle k'|\phi \rangle  \langle \phi | k  \rangle.
\label{eqn:def_VPS_BIC}
\end{equation}

\item[(iii)]
We allot a positive energy $E=K^2$ to $| \phi \rangle$,
to which  $V_0$ assigned  the negative energy $-\gamma_0^2$ 
in Eq.(\ref{eqn:Sch_bound_simple_method}).
Then, Eq.(\ref{eqn:def_VPB2}) immediately expresses the desired $V_{\rm B}$
as 
\begin{align}
\langle k' |V_{\rm B} | k  \rangle
&=  (K^2 - {k'}^2) 
     \langle k'|\phi \rangle  \langle \phi | k  \rangle
\nonumber\\
&=  (-\gamma_0^2 - {k'}^2) 
     \langle k'|\phi \rangle  \langle \phi | k  \rangle
+  (K^2 + {\gamma_0}^2) 
     \langle k'|\phi \rangle  \langle \phi | k  \rangle
\nonumber\\
&= \langle k' |V_{0,{\rm B}} | k  \rangle + \langle k' |\Delta V_{\rm B} | k  \rangle .
\label{eqn:def_VPB_BIC}
\end{align}
where we have defined 
\begin{align}
\Delta V_{\rm B} 
&\equiv V_{\rm B} - V_{0,{\rm B}}   
 =  (K^2 + \gamma_0^2) | \phi \rangle   \langle \phi | .
\label{eqn:expression_Delta_VB}
\end{align}
Note that $\Delta V_{\rm B}$ 
of Eq.(\ref{eqn:expression_Delta_VB}) is an energy-shift operator, i.e., 
it shifts the  eigenenergy $-\gamma_0^2$  of $| \phi \rangle$ to $K^2$.   
To be precise,  $H_0 + V =  H_0 + V_0 + \Delta V_{\rm B}$ has the same set of eigenstates
$\{| k \rangle\!_+,\,  |\phi \rangle\}$ as $H_0 + V_0$, 
but assigns a shifted eigenenergy
$K^2$ to $| \phi \rangle$.

\end{enumerate}
\noindent
With  $V_{\rm S}$ of Eq.(\ref{eqn:def_VPS_BIC}) and 
$V_{\rm B}$ of Eq.(\ref{eqn:def_VPB_BIC}), we define  
 $V$ by Eq.(\ref{eqn:V_eq_VPB_plus_VPS}), which gives
\begin{align}
\langle k' |V | k  \rangle 
&= \langle k' |V_0 | k  \rangle + \langle k' |\Delta V_{\rm B} | k  \rangle
\nonumber\\
&=
\langle k' |V_0 | k  \rangle + 
  (K^2 + \gamma_0^2) 
     \langle k'|\phi \rangle  \langle \phi | k  \rangle.
\label{eqn:def_V_single_BIC}
\end{align}
It is then easy to confirm 
  Schr\"odinger equation
(\ref{eqn:Sch_BIC_formal})
 with $\langle k' |V | k  \rangle$ of Eq.(\ref{eqn:def_V_single_BIC}),
which now reads
\begin{equation}
k^2 \phi(k) +   \int \!\! \frac{p^2 {\rm d}p}{(2\pi)^3} ~\langle k | V  | p  \rangle  \phi(p)
= K^2 \phi(k),
\label{eqn:Sch_BIC}
\end{equation}
in the same way 
as Eq.(\ref{eqn:Sch_bound}) in the general theory.
To summarize, 
Eq.(\ref{eqn:def_V_single_BIC}) defines the desired transformation
$V_0 \to V=V_0 + \Delta V_{\rm B}$, where $V$
describes the bound state  
$| \phi \rangle$ as a BIC at the positive energy $E=K^2$.\footnote{
The present method can be used also
to change the eigenenergy $-\gamma_0^2$ of $|\phi \rangle$ to another negative 
 value $-{\gamma_1}^2$; 
we just need to replace $K^2$ with $-{\gamma_1}^2$ in 
Eq.(\ref{eqn:def_V_single_BIC}).
} 
It is straightforward 
to generalize the above method to the most general case  with several  
bound states, some of which are at positive energies 
and others at negative energies.

The present transformation    $V_0 \to V=V_0 + \Delta V_{\rm B}$  
 is ``minimal'' in the sense that it only changes the eigenenergy 
of $|\phi \rangle$ as $-\gamma_0^2 \to K^2$, 
while leaving all the eigenstates,  $|\phi \rangle$ and $\{| k \rangle\!_+ \}$, untouched.    
Correspondingly, the SB-decomposition of the transformation 
is summarized by  $V_{0,{\rm B}} \to V_{\rm B}= V_{0,{\rm B}}+\Delta V_{\rm B}$ and 
$V_{\rm S} = V_{0,{\rm S}}$.
We say also that the present method is ``simple''
because we can immediately  
write down  the right hand side of Eq.(\ref{eqn:def_V_single_BIC})
 without complicated numerical calculations.  
So follows  the title of the present Sect.\,\ref{sec:simple_form_V_supporting_BIC}.  

\subsubsection{Comment}
\label{sec:simple_form_V_supporting_BIC_2}

If our goal were just to derive  Eq.(\ref{eqn:def_V_single_BIC}), 
the easiest way would be the following.
By definition,  we can express $H_0 + V_0$ and  $H_0 + V$ as
\begin{align}
H_0 + V_0 &~=~ 
\int \!\! \frac{k^2 {\rm d}k}{(2\pi)^3}~ k^2| k \rangle\!_+ {}_{+}\!\langle k | 
   ~ - ~ \gamma_0^2 | \phi \rangle   \langle \phi | ,
\label{eqn:expression_H0_plus_V0}
\\
H_0 + V &~=~ 
\int \!\! \frac{k^2 {\rm d}k}{(2\pi)^3}~ k^2| k \rangle\!_+ {}_{+}\!\langle k | 
   ~+~K^2 | \phi \rangle   \langle \phi | .
\label{eqn:expression_H0_plus_V}
\end{align}
Then, the difference  between Eqs.(\ref{eqn:expression_H0_plus_V})
and (\ref{eqn:expression_H0_plus_V0}) gives
\begin{align}
V -V_0 &=  (K^2 + \gamma_0^2) | \phi \rangle   \langle \phi | , 
\label{eqn:expression_V_minus_V0}
\end{align}
which  immediately leads to Eq.(\ref{eqn:def_V_single_BIC}).
However, the above derivation 
 does not SB-decompose $V$ of  Eq.(\ref{eqn:def_V_single_BIC})
into  $V_{\rm S}$ of Eq.(\ref{eqn:def_VPS_BIC}) and 
$V_{\rm B}$ of Eq.(\ref{eqn:def_VPB_BIC}). 
It would not allow, therefore, the detailed study 
 of $\langle k' |V | k  \rangle$  in the following sections, 
where we will make full use of the SB-decomposition.

\section{Numerical example I}
\label{sec:numerical_example_simple}

In this section, we explicitly construct a Hermitian potential supporting a BIC.
We use the simple method 
 in Sect.\,\ref{sec:simple_form_V_supporting_BIC} rather than 
the general method in Sect.\,\ref{sec:general_form_V_supporting_BIC},
because the minimal nature of its transformation $V_0 \to V=V_0 + \Delta V_{\rm B}$ is  
suitable  to  vary  
the eigenenergy of a fixed bound state  continuously  without changing scattering states.
In the following, we work in the S-wave channel for simplicity.

\subsection{Step (i) -- Gaussian potential $V_0$}
\label{sec:Gaussian_potential}
\begin{figure}[!htb]
\begin{center}
\vspace{-3mm}
\includegraphics[width=0.30\columnwidth]{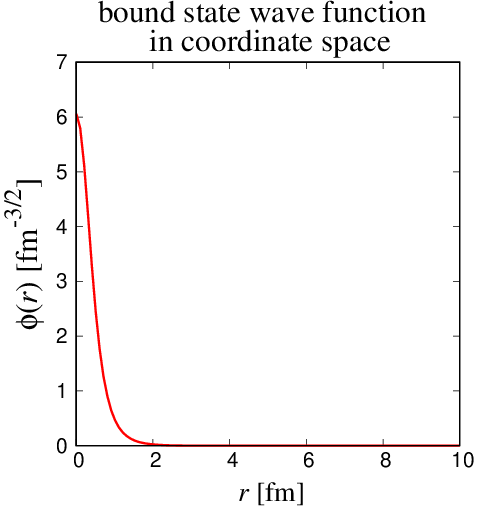}
\hspace{6mm}
\includegraphics[width=0.30\columnwidth]{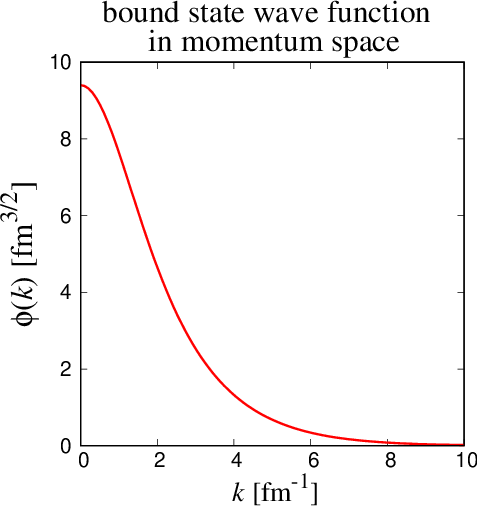}
\caption{
Wave function $\phi(r)$ of the bound state at 
   $-\gamma_0^2 = -5.373\, {\rm fm}^{-2} \,(= -222.8\, {\rm MeV})$
in coordinate space  (left panel), and $\phi(k)$ 
in  momentum space  (right panel)
of the Gaussian potential in Eq.(\ref{eqn:gauss_r}).
 See the text.}
\label{fig:bound_state_Gauss}
\end{center}
\end{figure}

\begin{figure}[!htb]
\begin{center}
\includegraphics[width=0.33\columnwidth]{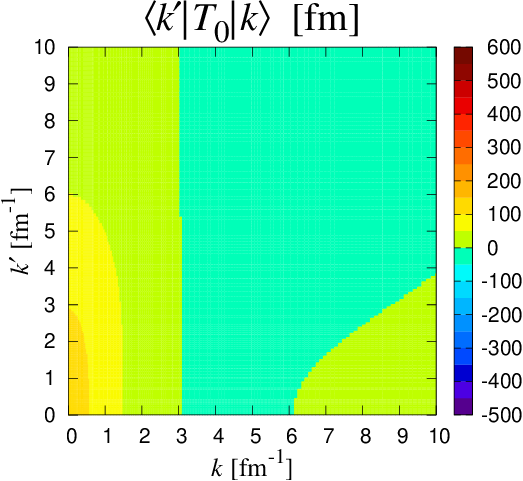}
\hspace{3mm}
\includegraphics[width=0.29\columnwidth]{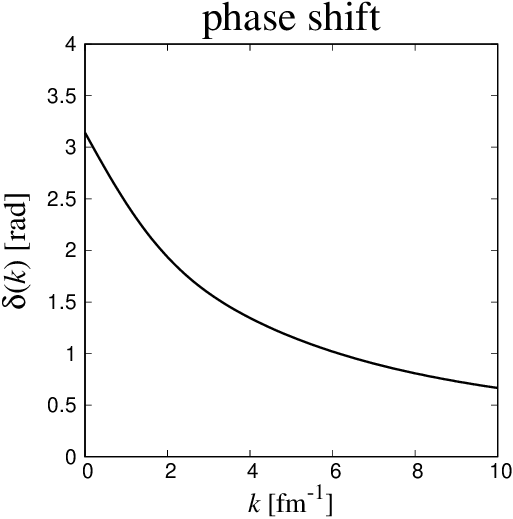}
\caption{
Half-on-shell T-matrix  $\langle k' |T_0| k  \rangle$ 
(left panel, contour map of the real part) and  phase shift 
$\delta(k)$  (right panel) of the Gaussian potential $V_0$ in Eq.(\ref{eqn:gauss_r}).
 See the text.}
\label{fig:delta_T_Gauss}
\end{center}
\end{figure}

Here, 
we adopt  the following local Gaussian potential for $V_0$ in step (i)
of  Sect.\,\ref{sec:simple_form_V_supporting_BIC}:\footnote{
In the calculations below, $m$
stands for the proton-neutron reduced mass, $ M_p  M_n / (M_p + M_n)$.
Then, our unit system, $\hbar=2m=1$, 
 gives $1=\hbar^2/(2m)=41.47{\,\rm MeV \,  fm^2}$.
Therefore, $\lambda=-30.0 \,{\rm fm}^{-2}$ corresponds to 
$-30.0 \,{\rm fm}^{-2}\times 41.47{\,\rm MeV  \, fm^2}
= -1244\,{\rm MeV}$.
}
\begin{equation}
V_0(r)
= \lambda \, \exp \!{\left(-\frac{r^2}{b^2} \right)} ,
~~~\lambda=-30.0 \,{\rm fm}^{-2},
~~b=0.5 \, {\rm fm},
\label{eqn:gauss_r}
\end{equation}
which is given in momentum space as
\begin{equation}
\langle k'| V_0| k \rangle
= 4\pi \lambda \, ({ b \sqrt{ \pi} })^3 
 \, j_0\left(-i \frac{kk'b^2}{2}\right)  \, \exp \! \left( -\dfrac{(k^2+{k'}^2)b^2}{4} \right),
\label{eqn:gauss_k}
\end{equation}
where $j_0(x)=\dfrac{\sin x}{x}$ is the spherical  Bessel function of zeroth order.
It is straightforward to show numerically that
the above $V_0$  supports a single bound state $|\phi \rangle$
at $ -\gamma_0^2 = -5.373\, {\rm fm}^{-2} \,(= -222.8\, {\rm MeV})$, 
of which the wave function
 $\phi(r)$  in coordinate space and $\phi(k)$ in momentum space 
are respectively shown in the left and the right panels 
in Fig.\ref{fig:bound_state_Gauss}.
For later convenience,  we also show  in Fig.\ref{fig:delta_T_Gauss}
 the half-on-shell T-matrix $\langle k' |T_0| k  \rangle$ and the phase shift $\delta(k)$ 
of $V_0$.

\vspace{-8mm}

\subsection{Step (ii) -- $V_{\rm S} = V_{0,{\rm S}}$}
\label{sec:Gaussian_potential_SB}

\begin{figure}[!htb]
\vspace{5mm}
\threeDGraphs{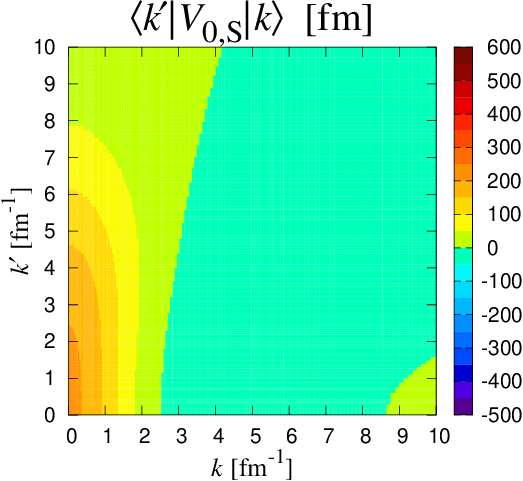}{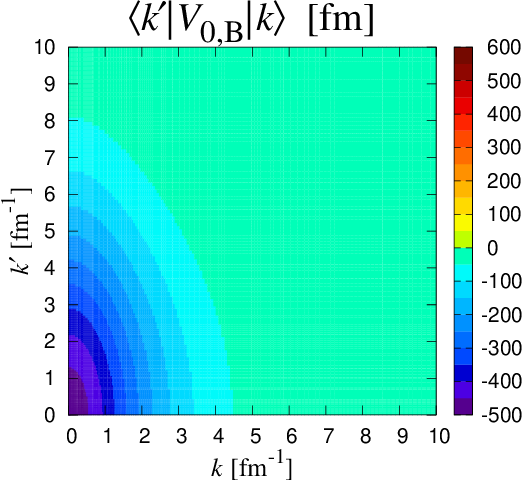}{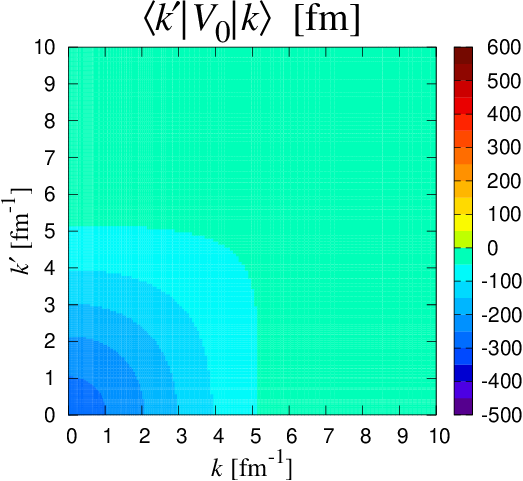}
\caption{
SB-decomposition $V_{0,{\rm S}} + V_{0,{\rm B}} = V_0$
 of the Gaussian potential $V_0$ 
in Eq.(\ref{eqn:gauss_k}) which has the bound state
  at $E=-\gamma_0^2 = -5.373\, {\rm fm}^{-2} \,(= -222.8\, {\rm MeV})$.
Left: contour map of $\langle k' |V_{0,{\rm S}}| k  \rangle$.
Middle: contour map of $\langle k' |V_{0,{\rm B}} | k  \rangle$.
Right: contour map of
 $\langle k' |V_0 | k  \rangle 
= \langle k' |V_{0,{\rm S}} | k  \rangle + \langle k' |V_{0,{\rm B}} | k  \rangle$.
 See the text.}
 \vspace{-5mm}
\label{fig:inv_example_init}
\end{figure}

Now 
we make the  SB-decomposition
of  $\langle k'| V_0| k \rangle$ in Eq.(\ref{eqn:gauss_k}),
to arrive at the results shown  
in Fig.\ref{fig:inv_example_init}.
In the right panel, we display 
$\langle k'| V_0| k \rangle$ of Eq.(\ref{eqn:gauss_k}),
which is decomposed into 
 $\langle k' |V_{0,{\rm B}} | k  \rangle$
of Eq.(\ref{eqn:def_V0PB_single_BIC}) and 
$\langle k' |V_{0,{\rm S}} | k  \rangle 
= \langle k' |V_{0} | k  \rangle - \langle k' |V_{0,{\rm B}} | k  \rangle$ 
shown in the middle and left panels, respectively.\footnote{
In the general method in Sect. \ref{sec:general_form_V_supporting_BIC},
we would have to calculate  $\langle k' |V_{0,{\rm S}} | k  \rangle$
via Eq.(\ref{eqn:def_VPS}) in terms of 
$\langle k' |T_{0} | k  \rangle$ in Fig.\ref{fig:delta_T_Gauss}.
 In the present ``simple'' method 
 in Sect.\,\ref{sec:simple_form_V_supporting_BIC},  on the other hand, 
we can immediately write down  $\langle k' |V_{0,{\rm S}} | k  \rangle$
without the knowledge of  $\langle k' |T_{0} | k  \rangle$.
}
Then,  following step (ii)  in Sect.\,\ref{sec:simple_form_V_supporting_BIC}, 
we define $V_{\rm S}$ by  
$\langle k'| V_{\rm S}| k \rangle=\langle k'| V_{0,{\rm S}}| k \rangle$.

We now examine Fig.\ref{fig:inv_example_init} before going to step (iii).  
First, let us look into $\langle k' |V_{0,{\rm B}} | k  \rangle$ 
in the middle panel.
Then,  Eq.(\ref{eqn:def_V0PB_single_BIC})
explains its behavior as follows;  
the  $k$-dependence of $\langle k' |V_{0,{\rm B}} | k  \rangle$ is 
given by $\phi (k)$ in Fig.\ref{fig:bound_state_Gauss}. 
On the other hand, its $k'$-dependence is given by 
$ (-\gamma_0^2 - {k'}^2) \phi( k')$,
which is negative and extends to higher momentum region than  
$ \phi(k)$.   Thus the product of these two factors
clearly explains $\langle k' |V_{0,{\rm B}} | k  \rangle$ in the middle panel.
Second, let us turn to $\langle k' |V_{0,{\rm S}} | k  \rangle$ 
in the left panel of Fig.\ref{fig:inv_example_init}.
We can see that, because the range of 
$\phi(k)$ is approximately $\sim \dfrac{1}{b} = 2.0 \, {\rm fm}^{-1}$,
$\langle k' |V_{0,{\rm S}} | k  \rangle$ is highly repulsive for $k \lesssim 2.0 \, {\rm fm}^{-1}$
to guarantee the orthogonality of $\{ | k \rangle\!_+ ,\, k \lesssim 2.0 \, {\rm fm}^{-1}\}$
to $| \phi \rangle$.
We also realize that, because $V_{0,{\rm S}}|\phi \rangle  = 0$ by definition,
$\langle k' |V_{0,{\rm S}} | k  \rangle$ becomes slightly  negative in the large-$k$ region
for each value of $k'$
 to guarantee 
${\displaystyle 
\langle k' |V_{0,{\rm S}}|\phi \rangle
= \int \!\! \frac{k^2 {\rm d}k}{(2\pi)^3}~ \langle k' |V_{0,{\rm S}}| k \rangle \, \phi(k)=0
}$.
The above observations explain main features of 
$\langle k' |V_{0,{\rm B}} | k  \rangle$ and 
$\langle k' |V_{0,{\rm S}} | k  \rangle$, and therefore of $\langle k' |V_{0} | k  \rangle$,
shown in Fig.\ref{fig:inv_example_init}.

\subsection{Step (iii) -- transformation $V_{0,{\rm B}} \to V_{\rm B}$}
\label{sec:transform_VB}

Now, we implement step (iii) of Sect.\,\ref{sec:simple_form_V_supporting_BIC}, i.e., 
we transform $\langle k' |V_{0,{\rm B}} | k  \rangle$
 of Eq.(\ref{eqn:def_V0PB_single_BIC})
into $\langle k' |V_{\rm B} | k  \rangle$ 
 of Eq.(\ref{eqn:def_VPB_BIC}).
For later convenience,
we replace  $K^2$ in $\langle k' |V_{\rm B} | k  \rangle$ 
 of Eq.(\ref{eqn:def_VPB_BIC}) with the energy variable $E$,
to express $\langle k' |V_{\rm B} | k  \rangle$ as a function of $E$ as
\begin{equation}
\langle k' |V_{\rm B} | k  \rangle
=  (E - {k'}^2) 
     \langle k'|\phi \rangle  \langle \phi | k  \rangle .
\label{eqn:def_VPB_single_E}
\end{equation}

\subsection{Numerical results for  $V_{\rm S} + V_{\rm B} = V$ in momentum space}
\label{sec:results_VS_plus_VB_makes_V}

\begin{figure}[!htb]
\vspace{10mm}
\threeDGraphs
 {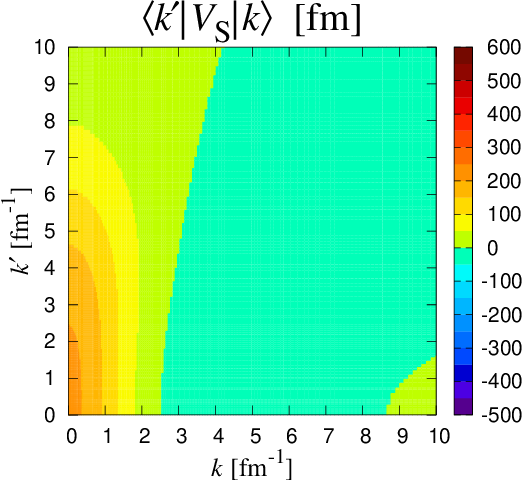}
 {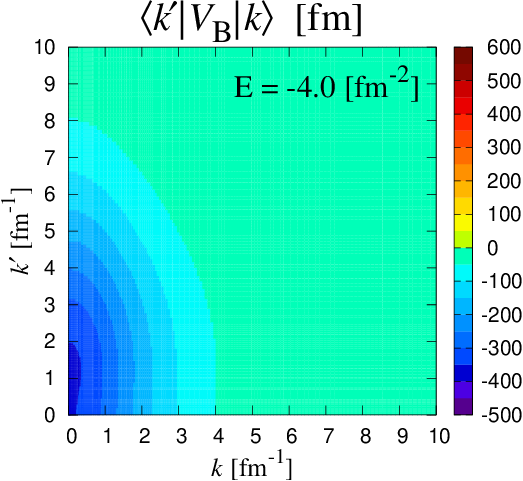}
 {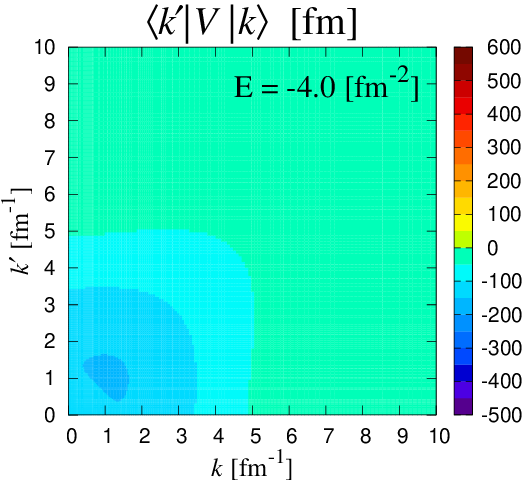}
\vspace{-5mm}
\caption{
SB-decomposition $V_{\rm S} + V_{\rm B} = V$ of $V$
 which gives the same $\langle k' |T| k  \rangle$
and $| \phi \rangle$ as  
those given by $\langle k' |V_{0}| k  \rangle$ of Fig.\ref{fig:inv_example_init}.
$\langle k' |V_{\rm S}| k  \rangle$ is the same as 
$\langle k' |V_{0,{\rm S}}| k  \rangle$ in Fig.\ref{fig:inv_example_init}.
In the expression (\ref{eqn:def_VPB_single_E}) for $\langle k' |V_{\rm B}| k  \rangle$, 
the eigenenergy $E$ of   $| \phi \rangle$ is chosen to be 
$E = -4.0\, {\rm fm}^{-2} \,(= -165.9\, {\rm MeV})$.
Other notation is the same as for Fig.\ref{fig:inv_example_init}.
 See the text.}
\label{fig:inv_example_4}
\end{figure}

\begin{figure}[!htb]
\vspace{10mm}
\threeDGraphs
{VPS_init_gauss_gamma2=5.3730.eps}
 {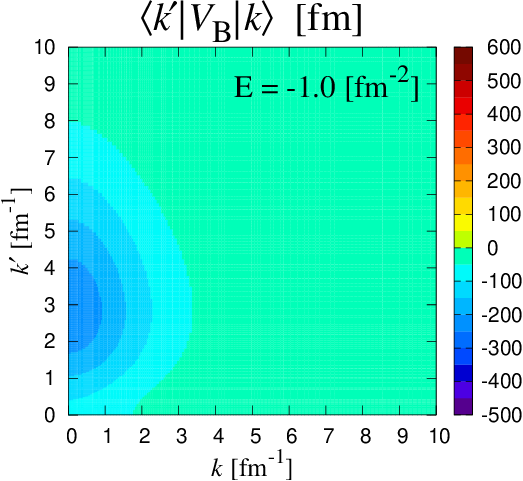}
 {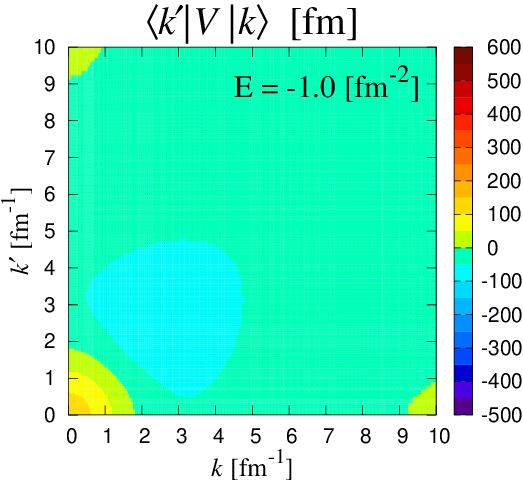}
\vspace{-5mm}
\caption{
SB-decomposition $V_{\rm S} + V_{\rm B} = V$. 
The eigenenergy of  $| \phi \rangle$ is 
$E = -1.0\, {\rm fm}^{-2} \,(= 41.47\, {\rm MeV})$.
Other notation is the same as for Fig.\ref{fig:inv_example_4}.
 See the text.}
\label{fig:inv_example_1}
\end{figure}

\begin{figure}[!htb]
\vspace{10mm}
\threeDGraphs
{VPS_init_gauss_gamma2=5.3730.eps}
 {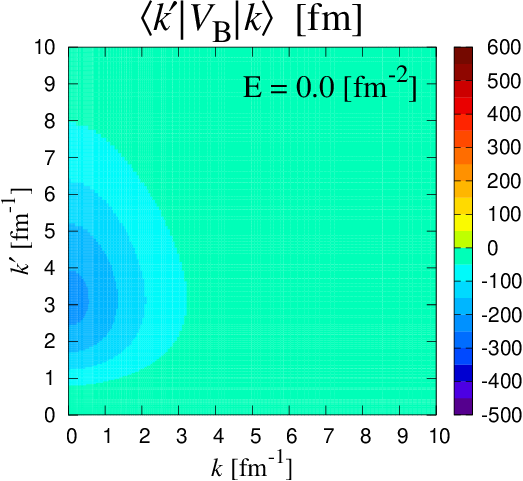}
 {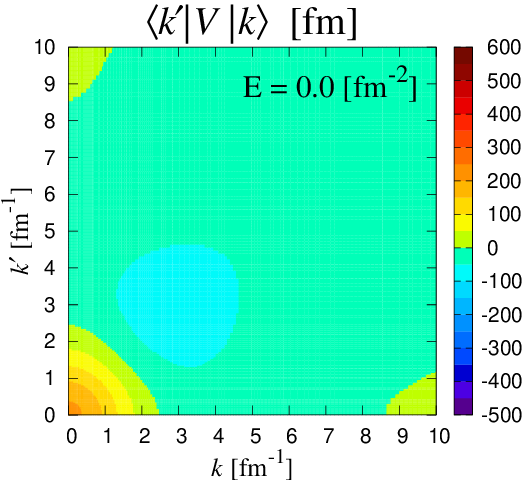}
\vspace{-5mm}
\caption{
SB-decomposition $V_{\rm S} + V_{\rm B} = V$.  
 The eigenenergy of  $| \phi \rangle$ is 
$E= 0.0\, {\rm fm}^{-2} \,(= 0.0\, {\rm MeV})$, i.e., $| \phi \rangle$
is a zero energy bound state.
Other notation is the same as for Fig.\ref{fig:inv_example_4}.
 See the text.}
\label{fig:inv_example_0}
\end{figure}

\begin{figure}[!htb]
\vspace{10mm}
\threeDGraphs
 {VPS_init_gauss_gamma2=5.3730.eps}
 {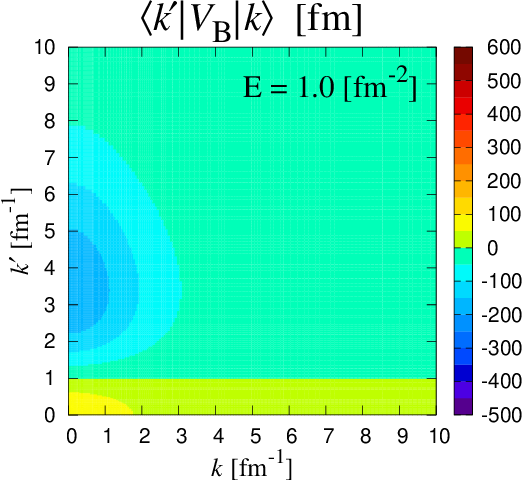}
{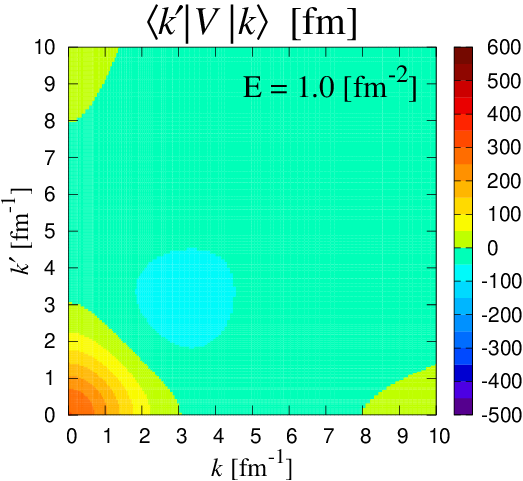}
\vspace{-5mm}
\caption{
SB-decomposition $V_{\rm S} + V_{\rm B} = V$.  
 The (positive) eigenenergy of  $| \phi \rangle$ is 
$E=K^2 = 1.0\, {\rm fm}^{-2} \,(= 41.47\, {\rm MeV})$.
Other notation is the same as for Fig.\ref{fig:inv_example_4}.
 See the text.}
  \vspace{-3mm}
\label{fig:inv_example_minus_1}
\end{figure}

\begin{figure}[!htb]
\vspace{10mm}
\threeDGraphs
 {VPS_init_gauss_gamma2=5.3730.eps}
 {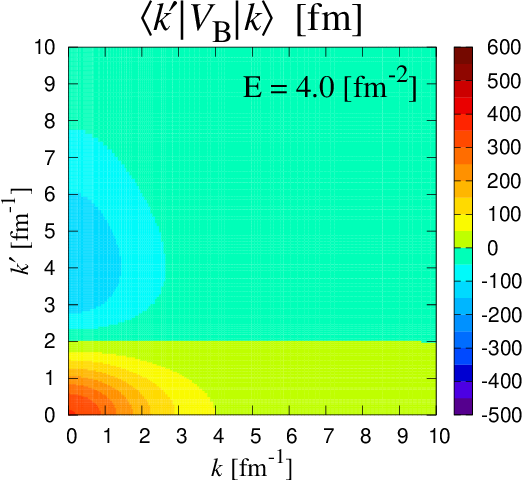}
 {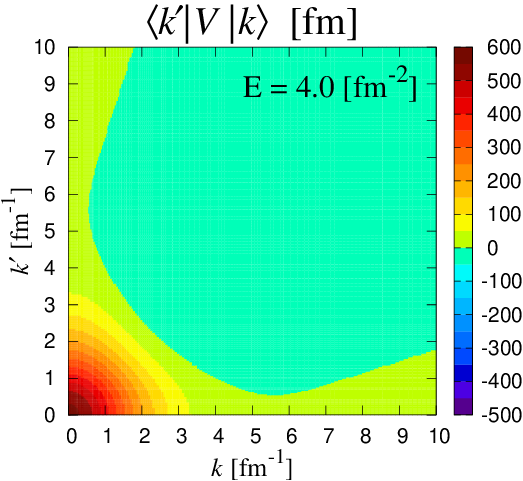}
\vspace{-5mm}
\caption{
SB-decomposition $V_{\rm S} + V_{\rm B} = V$. 
 The (positive) eigenenergy of  $| \phi \rangle$ is 
$E=K^2 = 4.0\, {\rm fm}^{-2} \,(= 165.9\, {\rm MeV})$.
Other notation is the same as for Fig.\ref{fig:inv_example_4}.
 See the text.}
 \vspace{-5mm}
\label{fig:inv_example_minus_4}
\end{figure}

We are now ready to investigate the transformed potential in Eq.(\ref{eqn:def_V_single_BIC}).
For each value of eigenenergy  $E$ of 
the fixed  bound state $|\phi \rangle$,  
$\langle k' |V_{\rm B} | k  \rangle$
of Eq.(\ref{eqn:def_VPB_single_E}), together with 
 $\langle k' |V_{\rm S} | k  \rangle =  \langle k' |V_{0,{\rm S}} | k  \rangle$
of Eq.(\ref{eqn:def_VPS_BIC}), 
makes its corresponding 
$\langle k' |V_{\rm S} | k  \rangle+ \langle k' |V_{\rm B} | k  \rangle 
= \langle k' |V | k  \rangle$, which is expressed in the form  of Eq.(\ref{eqn:def_V_single_BIC}) as
\begin{equation}
\langle k' |V | k  \rangle
=   \langle k' |V_0 | k  \rangle +   (E + \gamma_0^2) \langle k'  | \phi \rangle   \langle \phi | k  \rangle ,
\label{eqn:expression_VE_minus_V0}
\end{equation}
where we have replaced $K^2$ with $E$ as in Eq.(\ref{eqn:def_VPB_single_E}).\footnote{
$\langle k' |V | k  \rangle$ of Eq.(\ref{eqn:expression_VE_minus_V0}) reduces to 
$\langle k' |V_0 | k  \rangle$
Eq.(\ref{eqn:def_V_single_BIC}) for $E=-\gamma_0^2=-5.373 \,{\rm fm}^{-2}$.
}
Here we examine how the SB-decomposition 
 $ \langle k' |V_{0,{\rm S}} | k  \rangle + \langle k' |V_{0,{\rm B}} | k  \rangle
=\langle k' |V_0 | k  \rangle $ 
in Fig.\,\ref{fig:inv_example_init} changes 
 as we  continuously vary the eigenenergy  $E$. 
In actual calculations, we have varied 
$E=-\gamma_0^2=-5.373\, {\rm fm}^{-2}$ for $V_0$ (Fig.\ref{fig:inv_example_init}) 
to $ -4.0\, {\rm fm}^{-2}$ (Fig.\ref{fig:inv_example_4}), 
$-1.0\, {\rm fm}^{-2}$ (Fig.\ref{fig:inv_example_1}),
$0.0\, {\rm fm}^{-2}$ (Fig.\ref{fig:inv_example_0}),  
$1.0\, {\rm fm}^{-2}$ (Fig.\ref{fig:inv_example_minus_1}),
and  
$4.0\, {\rm fm}^{-2}$ (Fig.\ref{fig:inv_example_minus_4}), in order.\footnote{   
Note that  
each potential $\langle k' |V| k  \rangle$ in  
Figs.\,\ref{fig:inv_example_4} -- \ref{fig:inv_example_minus_4}
generally represents a  nonlocal potential  in coordinate space,
while  $\langle k' |V_0| k  \rangle$ 
in  Fig.\,\ref{fig:inv_example_init} stands for  
the local Gaussian potential $V_0(r)$ of  Eq.(\ref{eqn:gauss_r}).
}
We emphasize that  potentials in 
Figs.\,\ref{fig:inv_example_minus_1} and \ref{fig:inv_example_minus_4}
describe the bound state $| \phi \rangle$ as a BIC at positive energies
$E= 1.0\, {\rm fm}^{-2}$ and $4.0\, {\rm fm}^{-2}$, respectively.\footnote{
$\langle k' |V| k  \rangle$ in Fig.\,\ref{fig:inv_example_0} 
describes the bound state $| \phi \rangle$ at zero energy.
This might sound strange because 
it is  well known  that 
a local potential does not support a
zero energy bound state in the S-wave channel \cite{Newton82, Taylor83}.
Note, however, that the argument does not apply to  
nonlocal potentials \cite{Mulligan_76, Mulligan_81},
and that $\langle k' |V| k  \rangle$ in Fig.\,\ref{fig:inv_example_0}  is 
generally nonlocal as shall soon be shown by Eq.(\ref{eqn:transformed_gauss_r}).
}
In the following, we make several points 
using Figs.\,\ref{fig:inv_example_init} -- \ref{fig:inv_example_minus_4}.

First,  in going from  Fig.\,\ref{fig:inv_example_init}  to  Fig.\,\ref{fig:inv_example_minus_4},
we have only varied $\langle k' |V_{\rm B} | k  \rangle$ 
keeping $\langle k' |V_{\rm S}| k  \rangle = \langle k' |V_{0,{\rm S}}| k  \rangle$ untouched, i.e.,   
 we  have used the same $\langle k' | T_0  | k  \rangle$ 
 in Eq.(\ref{eqn:def_VPS})
for every $\langle k' |V_{\rm S} | k  \rangle$
 in  Figs.\,\ref{fig:inv_example_init} --  \ref{fig:inv_example_minus_4} 
as explained in step (ii) in Sect.\,\ref{sec:simple_form_V_supporting_BIC}.
Conversely, therefore, each 
$\langle k' |V | k  \rangle = \langle k' |V_{\rm S} | k  \rangle + \langle k' |V_{\rm B} | k  \rangle$ 
in  Figs.\,\ref{fig:inv_example_init} --  \ref{fig:inv_example_minus_4}
yields the same half-on-shell T-matrix 
$\langle k' | T  | k  \rangle = \langle k' | T_0  | k  \rangle$,
and therefore  the same phase shift $\delta(k)$.
Note also that 
$\langle k' |V | k  \rangle = \langle k' |V_{\rm S} | k  \rangle + \langle k' |V_{\rm B} | k  \rangle$
 in each figure is obviously Hermitian
as pointed in Sect.\,\ref{sec:general_design_V}.

Second, 
as $E$ increases from negative to positive values, it is visible that 
 the strong attraction of $\langle k' |V_{\rm B} | k  \rangle$ 
around the origin ($k=k'=0$) in Figs.\,\ref{fig:inv_example_init} and \ref{fig:inv_example_4}
gradually turns into the strong repulsion in Fig.\,\ref{fig:inv_example_minus_4}.
We can explain the above observation simply by 
noting that $\langle k' |V_{\rm B} | k  \rangle$ of Eq.(\ref{eqn:def_VPB_single_E}) 
can be approximated by  
$E  \langle 0|\phi \rangle  \langle \phi | 0  \rangle$ around the origin.
This in turn means that a positive $\langle 0 |V_{\rm B} | 0  \rangle$
is a clear indication of a BIC.\footnote{  
See also the discussion in Sect. \ref{sec:observing_BIC_1}.
}
Because $\langle k' |V_{\rm S} | k  \rangle$
is the same in  Figs.\,\ref{fig:inv_example_init} -- \ref{fig:inv_example_minus_4}, 
the above behavior of  $\langle k' |V_{\rm B} | k  \rangle$
also explains how $\langle k' |V| k  \rangle$ around the origin changes 
 with increasing $E$.

Third,  in order to examine  in detail  
 how the $k$ and $k'$-dependences of $\langle k' |V_{\rm B} | k  \rangle$ 
change with increasing $E$,
let us look into  Eq.(\ref{eqn:def_VPB_single_E}).
As in the discussion for $\langle k' |V_{0,{\rm B}} | k  \rangle$
in Sect.\,\ref{sec:Gaussian_potential_SB}, we can see that 
 the $k$-dependence of $\langle k' |V_{\rm B} | k  \rangle$ 
is given by $\phi (k)$ in Fig.\ref{fig:bound_state_Gauss}, 
which is independent of $E$ and is the same in 
Figs.\,\ref{fig:inv_example_init} --  \ref{fig:inv_example_minus_4}.
On the other hand, its $k'$-dependence
is given by $(E - {k'}^2)\phi (k')$ and varies with $E$.
For a negative energy $E= -\gamma^2$,   the factor 
$E - {k'}^2 = -\gamma^2 - {k'}^2$ is negative for any $k'$.
On the other hand, for  a positive energy $E=K^2$, 
 the factor $E - {k'}^2=K^2 - {k'}^2$
 changes its sign at $k' = K$, 
Therefore, because  $\phi (k)$ in Fig.\ref{fig:bound_state_Gauss} is nodeless, 
  $\langle k' |V_{\rm B} | k  \rangle$ of Eq.(\ref{eqn:def_VPB_single_E})  
changes its sign only once at $k' = K$
 as a function of $k'$; 
 it is visible that $\langle k' |V_{\rm B} | k  \rangle$ changes its 
sign at $k'=1.0 \,{\rm fm}^{-1}$ for each $k$  in  Fig.\ref{fig:inv_example_minus_1}, 
and  at $k'=2.0 \,{\rm fm}^{-1}$ in  Fig.\ref{fig:inv_example_minus_4}.
We stress that the above sign change of $\langle k' |V_{\rm B} | k  \rangle$
is a clear criterion of a BIC.

Fourth, let us compare  the potentials 
 in  Figs.\,\ref{fig:inv_example_1} and \ref{fig:inv_example_minus_1},
which describe the same state $|\phi \rangle$ 
as a negative energy eigenstate ($E= -1.0\, {\rm fm}^{-2}$) 
and a BIC  ($E= 1.0\, {\rm fm}^{-2}$), respectively.
 If we look into $\langle k' |V | k  \rangle$'s only,
we cannot realize that one of them supports a BIC while the other does not.
Only by the investigation of $\langle k' |V_{\rm B} | k  \rangle$ as in the above,
we do understand whether a given potential supports a BIC or not.\footnote{
In the search for a BIC,   
we only have to look for a sign change of $\langle k' |V_{\rm B} | k  \rangle$ as
a function of $k'$.  
 We observe the sign change 
 in Fig.\ref{fig:inv_example_minus_1} at $k'=1.0 \,{\rm fm}^{-1}$,
while we do not  in Fig.\ref{fig:inv_example_1}.
We can conclude, therefore, that 
$\langle k' |V | k  \rangle$ in Fig.\ref{fig:inv_example_minus_1}
 supports a BIC at $E = {k'}^2 = 1.0 \,{\rm fm}^{-2}$, 
while $\langle k' |V | k  \rangle$ in Fig.\ref{fig:inv_example_1} does not.
In the above discussion,  we have assumed a system with a single bound state. 
For a general case with several bound states, see Sect.\,\ref{sec:observing_BIC_2}.
}

To summarize, by using the simple method
 in Sect.\,\ref{sec:simple_form_V_supporting_BIC},
we have explicitly constructed potentials that support a BIC, which are shown in
 Figs.\ref{fig:inv_example_minus_1} and \ref{fig:inv_example_minus_4}.
Further, we have explained  major features of these potentials
using the SB-decomposition in  Sect.\,\ref{sec:general_design_V},
and have shown a criterion to judge whether a given potential has  a BIC or not.

\subsection{Numerical results for  $V_{\rm S} + V_{\rm B} = V$ in coordinate space}
\label{sec:results_VS_plus_VB_makes_V_r}
In order to obtain a better insight into BIC-supporting potentials, 
 we now cast the results in Sect.\,\ref{sec:results_VS_plus_VB_makes_V} into coordinate space.    

First, let us note that the Fourier transform of 
$\langle k'| V_{0}| k \rangle$ of Eq.(\ref{eqn:gauss_k})
is given by 
\begin{align}
\langle r'| V_0| r \rangle &\, \equiv \, 
\int \!\! \frac{{k'}^2 {\rm d}k'}{(2\pi)^3}\,4\pi j_0(k'r')\,
\int \!\! \frac{{k}^2 {\rm d}k}{(2\pi)^3}\,4\pi j_0(kr)~\langle k'| V_0| k \rangle
\nonumber\\
& \, = \, 
\dfrac{\delta(r'-r)}{r'r} V_0(r),
\label{eqn:transformed_gauss_r_1}
\end{align}
where $V_0(r)$ is the Gaussian potential of Eq.(\ref{eqn:gauss_r}).
Second,   we can easily Fourier transform numerically 
$\langle k'| V_{\rm S}| k \rangle= \langle k'| V_{0,{\rm S}}| k \rangle$
 in Eq.(\ref{eqn:def_VPS_BIC})
and $\langle k'| V_{\rm B}| k \rangle$ in Eq.(\ref{eqn:def_VPB_BIC})
to obtain 
 $\langle r'| V_{\rm S}| r \rangle$ 
and  $\langle r'| V_{\rm B}| r \rangle$, which now read\footnote{
Here, we have replaced $K^2$ with $E$ as in Eq.(\ref{eqn:def_VPB_single_E}).
}
\begin{align}
 \langle r'| V_{\rm S}| r \rangle
&\,=\,\langle r'| V_{0,{\rm S}}| r \rangle \,=\,\langle r'| V_0| r \rangle - \langle r'| V_{0,{\rm B}}| r \rangle
\nonumber\\
&\,=\, \dfrac{\delta(r'-r)}{r'r} V_0(r) \, - \, 
\int \!\! \frac{{k'}^2 {\rm d}k'}{(2\pi)^3}\,4\pi j_0(k'r')
\int \!\! \frac{{k}^2 {\rm d}k}{(2\pi)^3}\,4\pi j_0(kr)~ (-\gamma_0^2-{k'}^2) \phi (k') \phi(k)
\nonumber\\
&\,=\, \dfrac{\delta(r'-r)}{r'r} V_0(r) \, - \, 
 (-\gamma_0^2+{\nabla'}^2) \phi (r') \phi(r).
\label{eqn:VS_r}
\\
 \langle r'| V_{\rm B}| r \rangle
&\,=\, 
\int \!\! \frac{{k'}^2 {\rm d}k'}{(2\pi)^3}\,4\pi j_0(k'r') 
\int \!\! \frac{{k}^2 {\rm d}k}{(2\pi)^3}\,4\pi j_0(kr)~(E-{k'}^2) \phi (k') \phi(k)
\nonumber\\
&\,=\,  (E+{\nabla'}^2) \phi (r') \phi(r).
\label{eqn:VB_r}
\end{align}
Here, we have used the notation 
$\nabla'^2 = \dfrac{{\rm d}^2}{{\rm d}r'^2} + \dfrac{2}{r'}\dfrac{{\rm d}}{{\rm d}r'}$.
Then,  by adding Eqs.(\ref{eqn:VS_r}) and (\ref{eqn:VB_r}), 
we can express the Fourier transform of
 Eq.(\ref{eqn:expression_VE_minus_V0}) as 
\begin{align}
\langle r'| V_{\rm S}| r \rangle + \langle r'| V_{\rm B}| r \rangle
&= 
\langle r'| V| r \rangle
\nonumber\\
&= \langle r'| V_0| r \rangle  - \langle r'| V_{0,{\rm B}}| r \rangle  + \langle r'| V_{\rm B}| r \rangle 
= \langle r'| V_0| r \rangle   + \langle r'| \Delta V_{\rm B}| r \rangle 
\nonumber\\
&= \dfrac{\delta(r'-r)}{r'r} V_0(r) \, + \, (E + \gamma_0^2) \phi (r') \phi(r).
\label{eqn:transformed_gauss_r}
\end{align}
The first line of Eq.(\ref{eqn:transformed_gauss_r}),
with $\langle r'| V_{\rm S}| r \rangle$ and  $\langle r'| V_{\rm B}| r \rangle$
being given by Eqs.(\ref{eqn:VS_r}) and (\ref{eqn:VB_r}) respectively, 
represents the SB-decomposition of $\langle r'| V| r \rangle$
which supports the bound state $\phi(r)$ at energy $E$.
The third line represents the Fourier transform of  Eq.(\ref{eqn:def_V_single_BIC}) 
for $\langle k'| V| k \rangle$, and gives the energy-shift operator $\Delta V_{\rm B}$
of Eq.(\ref{eqn:expression_Delta_VB}) in coordinate space:
\begin{align}
\langle r'| \Delta V_{\rm B}| r \rangle 
&= \langle r'| V_{\rm B} - V_{0,{\rm B}}| r \rangle
=(E + \gamma_0^2) \phi (r') \phi(r).
\label{eqn:shift_operator_delta_VB}
\end{align}
It is notable here that $\langle r'| V_{\rm B}| r \rangle$ of Eq.(\ref{eqn:VB_r})
is purely nonlocal and separable, 
and so is  $\langle r'| \Delta V_{\rm B}| r \rangle$ of Eq.(\ref{eqn:shift_operator_delta_VB}).   

\begin{figure}[!htb]
\vspace{6mm}
\threeDGraphs
{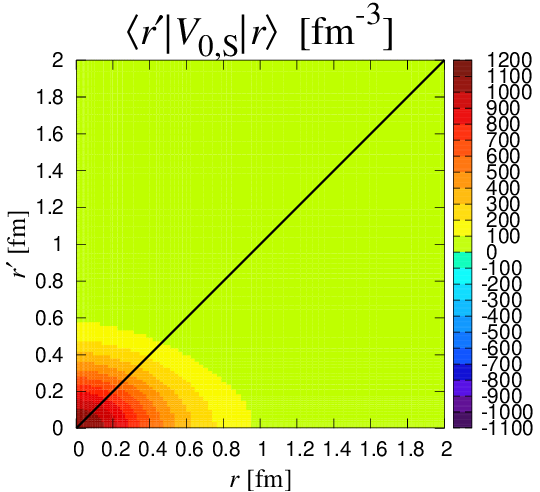}
{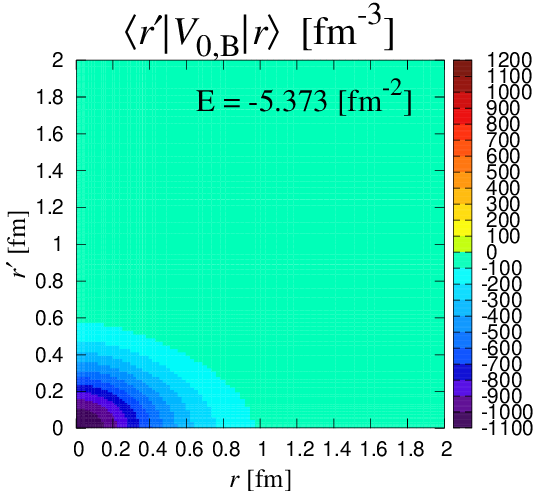}
{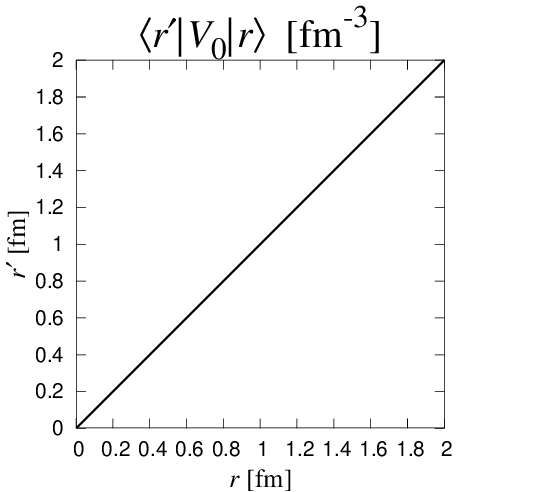}
\vspace{-3mm}
\caption{
SB-decomposition $V_{0,{\rm S}} + V_{0,{\rm B}} = V_0$
in the   first line of Eq.(\ref{eqn:transformed_gauss_r}) 
with the eigenenergy of $\phi(r)$ being 
$E=-\gamma_0^2 = -5.373\, {\rm fm}^{-2} \,(= -222.8\, {\rm MeV})$, 
which corresponds to Fig.\,\ref{fig:inv_example_init} in momentum space.
Left: $\langle r' |V_{\rm S}| r  \rangle = \langle r' |V_{0,{\rm S}}| r  \rangle$.
Middle: $\langle r' |V_{0,{\rm B}} | r  \rangle$.
Right: 
 $\langle r' |V_0 | r  \rangle = \dfrac{\delta(r'-r)}{r'r} V_0(r)$ which is 
shown by a thick black line along $r'=r$. 
Note the same thick black line  in the left panel.
 See the text.}
\label{fig:inv_example_init_r}
\end{figure}

\begin{figure}[!htb]
\vspace{5mm}
\threeDGraphs
{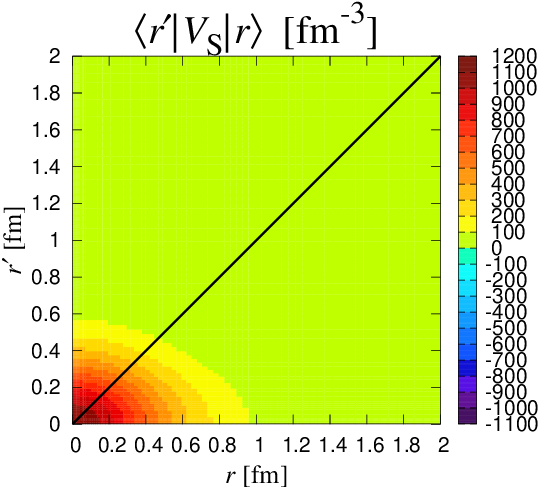}
{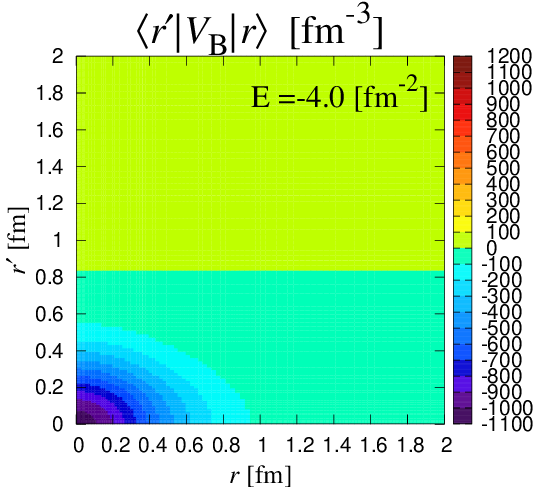}
{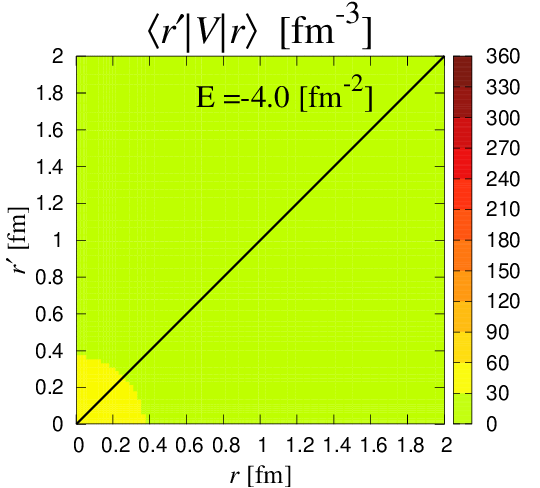}
\caption{
SB-decomposition $V_{\rm S} + V_{\rm B} = V$
 in coordinate space which corresponds to Fig.\,\ref{fig:inv_example_4}.
The eigenenergy of  $| \phi \rangle$ is 
$E = -4.0\, {\rm fm}^{-2}\,(= -165.9\, {\rm MeV})$.
The diagonal solid lines along $r'=r$ in the left and the right panels 
 represent the original local potential
$\langle r' |V_0 | r  \rangle = \dfrac{\delta(r'-r)}{r'r} V_0(r)$  in 
Fig.\ref{fig:inv_example_init_r}.
Other notation is the same as for Fig.\ref{fig:inv_example_init_r}.
 See the text.} 
\label{fig:inv_example_4_r}
\end{figure}

\begin{figure}[!htb]
\vspace{5mm}
\threeDGraphs
{Vsrr_fixed_phase.eps}
{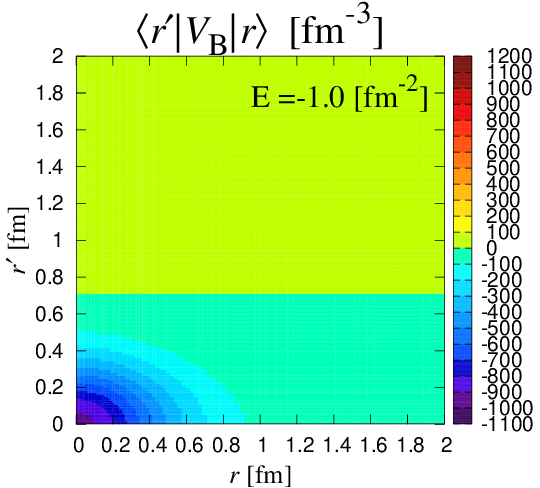}
{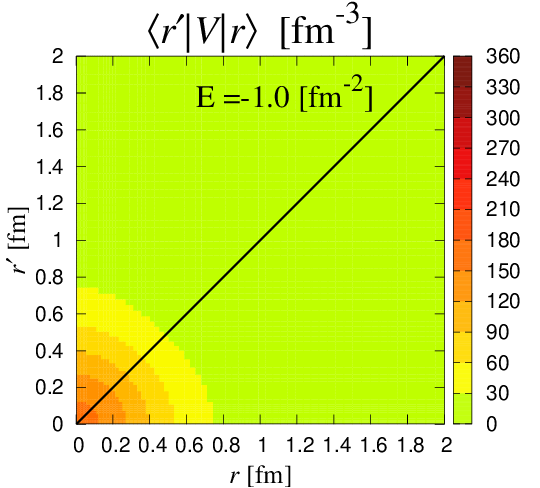}
\vspace{-5mm}
\caption{
SB-decomposition $V_{\rm S} + V_{\rm B} = V$
 in coordinate space which corresponds to Fig.\,\ref{fig:inv_example_1}.
The eigenenergy of  $| \phi \rangle$ is 
$E = -1.0\, {\rm fm}^{-2}\,(= -41.47\, {\rm MeV})$.
Other notation is the same as for Fig.\ref{fig:inv_example_init_r}.
 See the text.}
\label{fig:inv_example_1_r}
\end{figure}

\begin{figure}[!htb]
\vspace{10mm}
\threeDGraphs
{Vsrr_fixed_phase.eps}
{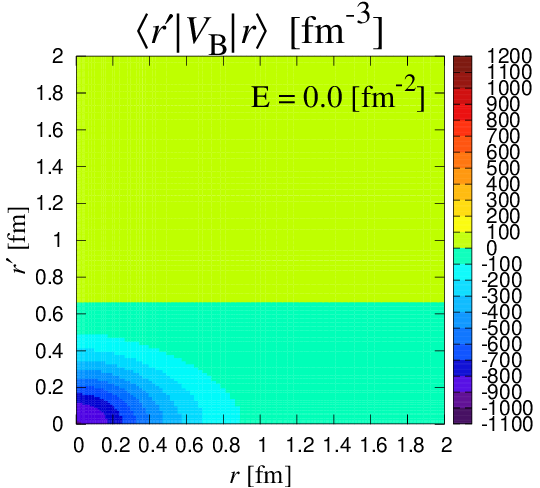}
{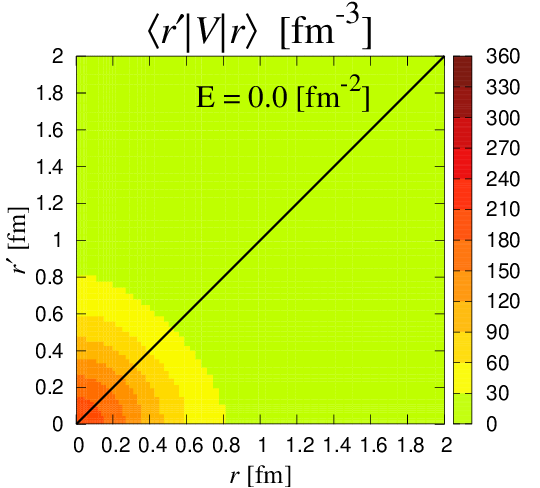}
\vspace{-5mm}
\caption{
SB-decomposition $V_{\rm S} + V_{\rm B} = V$
 in coordinate space which corresponds to Fig.\,\ref{fig:inv_example_0}.
The eigenenergy of  $| \phi \rangle$ is 
$E= 0.0\, {\rm fm}^{-2}\,(= 0.0\, {\rm MeV})$.
Other notation is the same as for Fig.\ref{fig:inv_example_init_r}.
 See the text.}
\label{fig:inv_example_0_r}
\end{figure}

\begin{figure}[!htb]
\vspace{10mm}
\threeDGraphs
{Vsrr_fixed_phase.eps}
{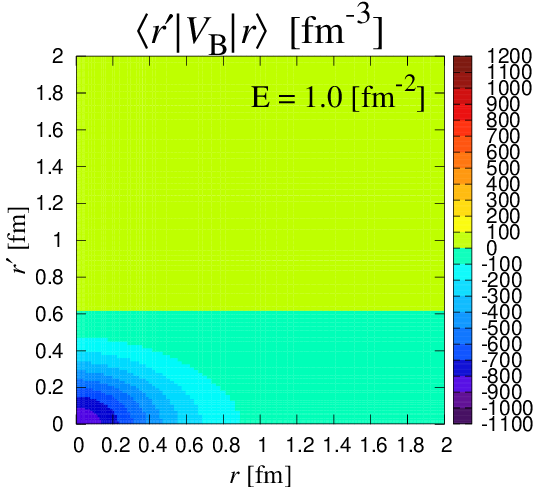}
{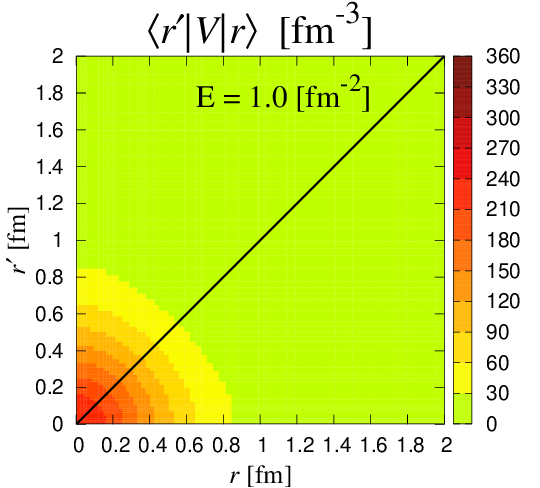}
\vspace{-5mm}
\caption{
SB-decomposition $V_{\rm S} + V_{\rm B} = V$
 in coordinate space which corresponds to Fig.\,\ref{fig:inv_example_minus_1}.
The eigenenergy of  $| \phi \rangle$ is 
$E=K ^2 = 1.0\, {\rm fm}^{-2}\,(= 41.47\, {\rm MeV})$.
Other notation is the same as for Fig.\ref{fig:inv_example_init_r}.
 See the text.}
\label{fig:inv_example_minus_1_r}
\end{figure}

\begin{figure}[!htb]
\vspace{10mm}
\threeDGraphs
{Vsrr_fixed_phase.eps}
{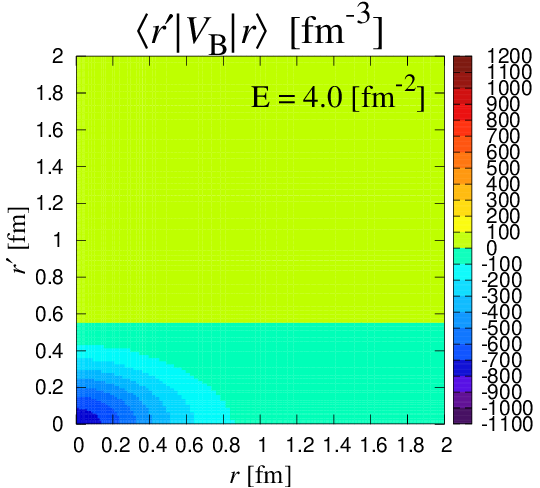}
{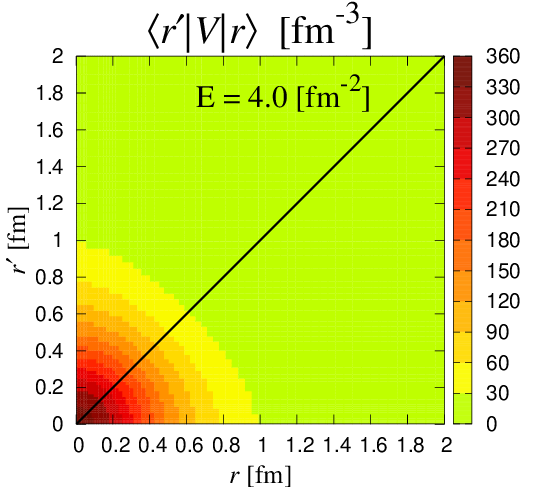}
\vspace{-5mm}
\caption{
SB-decomposition $V_{\rm S} + V_{\rm B} = V$
 in coordinate space which corresponds to Fig.\,\ref{fig:inv_example_minus_4}.
The eigenenergy of  $| \phi \rangle$ is 
$E=K ^2 = 4.0\, {\rm fm}^{-2}\,(= -165.9\, {\rm MeV})$.
Other notation is the same as for Fig.\ref{fig:inv_example_init_r}.
 See the text.}
\label{fig:inv_example_minus_4_r}
\end{figure}

In Figs.\,\ref{fig:inv_example_init_r} -- \ref{fig:inv_example_minus_4_r}, 
we present the first line of Eq.(\ref{eqn:transformed_gauss_r}),
 the SB-decomposition of $\langle r'| V| r \rangle$, corresponding to 
Figs.\,\ref{fig:inv_example_init} -- \ref{fig:inv_example_minus_4} 
in momentum space.

First of all,  the transformation  $V_0 \to V=V_0+\Delta V_{\rm B}$ 
in the last line of Eq.(\ref{eqn:transformed_gauss_r}) 
shows that the right panels in 
Figs.\,\ref{fig:inv_example_4_r} -- \ref{fig:inv_example_minus_4_r} 
represent, apart from the diagonal thick black lines representing 
$\langle r' |V_0 | r  \rangle = \dfrac{\delta(r'-r)}{r'r} V_0(r)$, 
the energy-shift operator 
$\langle r'| \Delta V_{\rm B}| r \rangle$
of Eq.(\ref{eqn:shift_operator_delta_VB}).
For $E = -\gamma_0^2$, $\langle r'| \Delta V_{\rm B}| r \rangle$ vanishes by definition, 
and $V$ of Eq.(\ref{eqn:transformed_gauss_r})
reduces to $V_0$,
to give 
\begin{equation}
\langle r'| V_{0,{\rm S}}| r \rangle + \langle r'| V_{0,{\rm B}}| r \rangle
= \dfrac{\delta(r'-r)}{r'r} V_0(r),
\label{eqn:transformed_gauss_0_r}
\end{equation}
showing that 
$\langle r'| V_{0,{\rm S}}| r \rangle$  and $\langle r'| V_{0,{\rm B}}| r \rangle$
cancel each other exactly for $r' \ne r$
as can be confirmed in Fig.\,\ref{fig:inv_example_init_r}.

\begin{figure}[!htb]
\begin{center}
\vspace{-3mm}
\includegraphics[width=0.35\columnwidth] {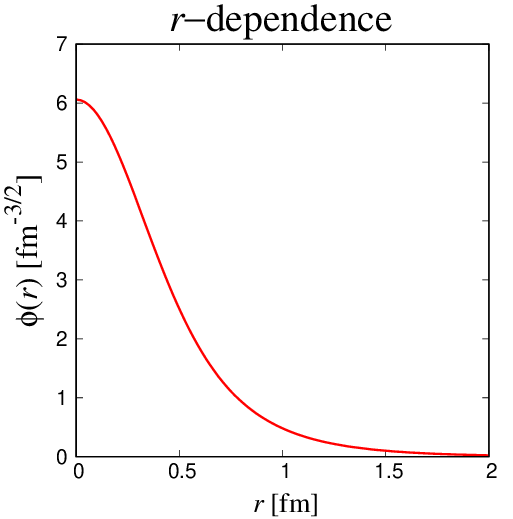}
\hspace{5mm}
\includegraphics[width=0.35\columnwidth]{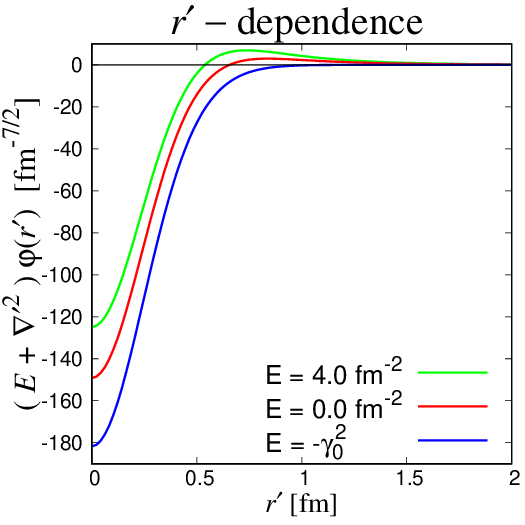}
\caption{
$r$ and $r'$-dependences of 
$\langle r' |V_{\rm S} | r  \rangle$ 
and $\langle r' |V_{\rm B} | r  \rangle$.   
Left:  $\phi (r)$ representing 
the $r$-dependence of 
both $\langle r' |V_{\rm S} | r  \rangle$  for $r' \ne r$ 
and $\langle r' |V_{\rm B} | r  \rangle$.
Right:  $(E+{\nabla'}^2) \phi (r')$  representing 
the $r'$-dependence of $\langle r' |V_{\rm B} | r  \rangle$
for several values of $E$
 $(4.0\, {\rm fm}^{-2},\,0.0\, {\rm fm}^{-2},\,-\gamma_0^2=-5.373\, {\rm fm}^{-2})$.
The curve $E=-\gamma_0^2$ gives (minus of) the $r'$-dependence of 
$\langle r' |V_{\rm S} | r  \rangle=\langle r' |V_{0,{\rm S}} | r  \rangle$.   
The curve for $E=0$ shows ${\nabla'}^2 \phi (r')$.
See the text.}
\label{fig:Fourier_phi_kkphi}
\end{center}
\end{figure}

Now, we examine the $r$ and $r'$-dependences of 
$\langle r' |V_{\rm S} | r  \rangle$ 
and $\langle r' |V_{\rm B} | r  \rangle$ in these figures.
For this purpose,
we plot $\phi (r)$ and $(E+{\nabla'}^2)\phi (r')$ respectively 
in the left and the right panels of Fig.\,\ref{fig:Fourier_phi_kkphi}.\footnote{
  $\phi (r)$ has been shown already in Fig.\,\ref{fig:bound_state_Gauss}.
For convenience, however, we plot it here again.
}
We understand the curves   in the right panel in the following manner.  
First, let us note that  ${\nabla'}^2 \phi (r')$, which is given by the curve $E=0$,  
 has a shorter range than $\phi (r)$ in the left panel, 
and has a node at $r'=0.67\, {\rm fm}$.\footnote{
Observe that ${\nabla'}^2 \phi (r')$ is the Fourier transform of 
$k'^2 \phi (k')$ in Eq.(\ref{eqn:VS_r}) 
 which has a longer rage in momentum space than $\phi (k')$.
Therefore, ${\nabla'}^2 \phi (r')$ has a shorter range than $\phi (r')$
in coordinate space.
Note also that ${\nabla'}^2 \phi (r')$,
 mostly the second derivative of $\phi (r')$ in the left panel, naturally has a node.
}
The node appears for 
$E>-\gamma_0^2$, and moves inward with increasing $E$.\footnote{ 
$\phi (r')$ decays as $\exp(-\gamma_0 r')$ for $r' \to \infty$, and so does ${\nabla'}^2 \phi (r')$.
Further, Schr\"odinger  equation 
for $\phi (r')$ tells that
$(-\gamma_0^2+{\nabla'}^2) \phi (r') = V_0(r')\phi (r')$ decays  faster 
than  $\exp(-\gamma_0 r')$, 
showing that leading (exponentially decaying) parts of 
$-\gamma_0^2 \phi (r')$ and ${\nabla'}^2 \phi (r')$ cancel each other exactly  
for $r' \to \infty$.
Therefore, $\phi (r)$ in the left panel and ${\nabla'}^2 \phi (r')$
in the right panel (indicated by $E=0$) explains that 
$(E+{\nabla'}^2) \phi (r')$  shows up with a node  for $E > -\gamma_0^2$ 
while it is nodeless for $E < -\gamma_0^2$.
}

Let us start with  $\langle r' |V_{\rm S} | r  \rangle$, 
which is independent of $E$ and therefore is common in 
Figs.\,\ref{fig:inv_example_init_r} -- \ref{fig:inv_example_minus_4_r}.  
For $r' \ne r$, $\langle r' |V_{\rm S} | r  \rangle$ reduces to  
$-\langle r' |V_{0,{\rm B}} | r  \rangle = (\gamma_0^2-{\nabla'}^2) \phi (r')  \phi (r)$,
 which is 
the second term of the right hand side of  Eq.(\ref{eqn:VS_r}).
Its $r$ and  $r'$-dependences are respectively given 
 by $\phi (r)$ in the left panel and 
by $(-\gamma_0^2+{\nabla'}^2) \phi (r')$
 in the right panel of Fig.\,\ref{fig:Fourier_phi_kkphi}. 
We see that  $\langle r' |V_{\rm S} | r  \rangle$ for $r' \ne r$  
has no node and that its extension is longer 
in the $r$-direction than in the $r'$-direction,
which is explained by the above observations
 on  Fig.\,\ref{fig:Fourier_phi_kkphi}.

Now we turn to  $\langle r' |V_{\rm B} | r  \rangle$ of  Eq.(\ref{eqn:VB_r}).
Its  $r$-dependence is independent of $E$, and is  given simply by $\phi (r)$ 
in the same way as $\langle r' |V_{\rm S} | r  \rangle$ for $r' \ne r$.
Its $r'$-dependence, on the other hand, is dependent on $E$, 
and is given by  $(E+{\nabla'}^2) \phi (r')$ in the right panel of Fig.\,\ref{fig:Fourier_phi_kkphi}; 
for $E > -\gamma_0^2$, it  has a node which moves inward as $E$ increases, i.e., 
 the node is located at  $r' \sim 0.67 \,{\rm fm}$ for $E= 0.0\, {\rm fm}^{-2}$,
and $r' \sim 0.54 \,{\rm fm}$ for $E= 4.0\, {\rm fm}^{-2}$.
We can thus  explain 
 the $r'$-dependence of $\langle r' |V_{\rm B} | r  \rangle$
in Figs.\,\ref{fig:inv_example_init_r} --  \ref{fig:inv_example_minus_4_r};
 $\langle r' |V_{\rm B} | r  \rangle$ has a node 
 in the $r'$-direction for $E > -\gamma_0^2$, which  is visible at  
$r' \sim 0.67 \,{\rm fm}$ in Fig.\,\ref{fig:inv_example_4_r} for $E= 0.0\, {\rm fm}^{-2}$,
 and goes downward as $E$ increases to $r' \sim 0.54 \,{\rm fm}$ in 
Fig.\,\ref{fig:inv_example_minus_4_r} for $E= -4.0\, {\rm fm}^{-2}$.

To summarize, we have understood main behaviors of 
$\langle r' |V_{\rm S} | r  \rangle$ and $\langle r' |V_{\rm B} | r  \rangle$, 
and therefore of $\langle r' |V| r  \rangle$
in Figs.\,\ref{fig:inv_example_init_r} --  \ref{fig:inv_example_minus_4_r}.
Note that the present results imply the importance of the nonlocal potential
in the physics of BIC; 
in order to realize a BIC in a
given system with a negative energy bound state, 
 it is crucial to introduce the nonlocal  perturbation
$\langle r'| \Delta V_{\rm B}| r \rangle$
to the system.   
We shall soon find a rigorous proof of this observation  
in Sect.\,\ref{sec:local_vs_nonlocal}.

\section{Numerical example II -- general method}
\label{sec:numerical_example_general}

In this section, we briefly  demonstrate 
that steps (i) -- (iii) of the general method in 
Sect.\,\ref{sec:general_form_V_supporting_BIC}
are really implementable. 

In  step (i) of Sect.\,\ref{sec:general_form_V_supporting_BIC},
we arbitrarily adopt the bound state $| \phi \rangle$ in Fig.\,\ref{fig:bound_state_Gauss}
for simplicity.\footnote{
Of course, we could have chosen a different bound state.
For the sake of comparison with the results in   Sect.\,\ref{sec:numerical_example_simple},
however, we have chosen the same bound state $| \phi \rangle$
as in Sect.\,\ref{sec:numerical_example_simple}.
We stress that we do not need knowledge of the Gaussian potential of Eq.(\ref{eqn:gauss_r})
in carrying out the general method here,
while we used it in implementing the simple method 
in Sect.\,\ref{sec:numerical_example_simple}.
}

\begin{figure}[!htb]
\begin{center}
\vspace{-3mm}
\includegraphics[width=0.33\columnwidth]{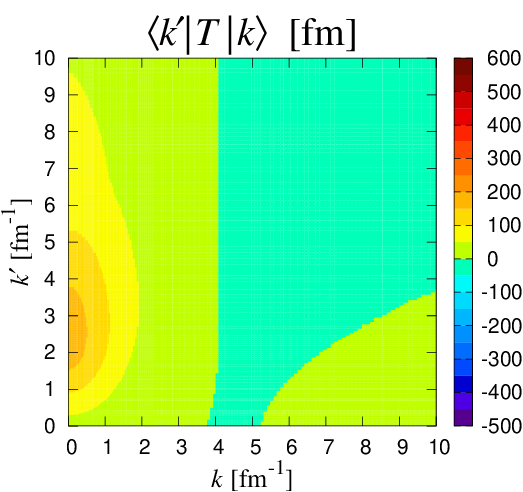}
\hspace{3mm}
\includegraphics[width=0.29\columnwidth]{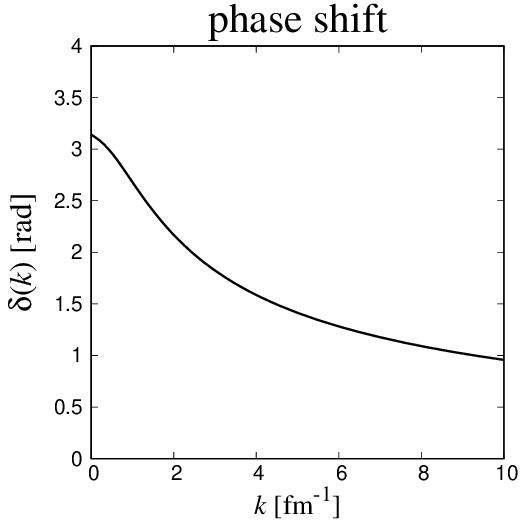}
\caption{
Arbitrarily chosen solution 
$\langle k' |T | k  \rangle$  to  Eqs.({\bf A}) and ({\bf B})
(contour map of the real part, left panel) and 
its corresponding phase shift $\delta(k)$ (right panel).
Notation is the same as for Fig.\ref{fig:delta_T_Gauss}.
 See the text.}
\label{fig:delta_T_general}
\end{center}
\end{figure}	

\begin{figure}[!htb]
\vspace{10mm}
\threeDGraphs
{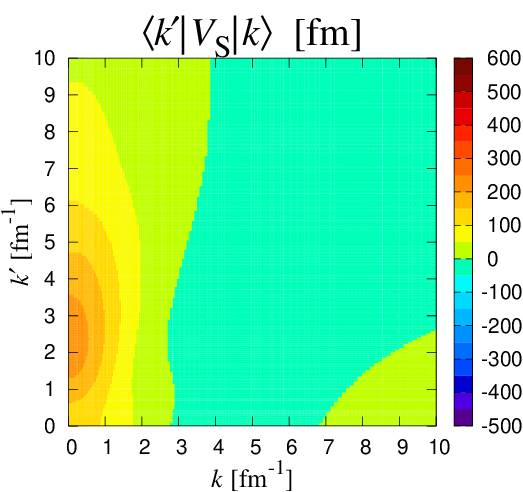}
 {VPB_gamma2=4.eps}
{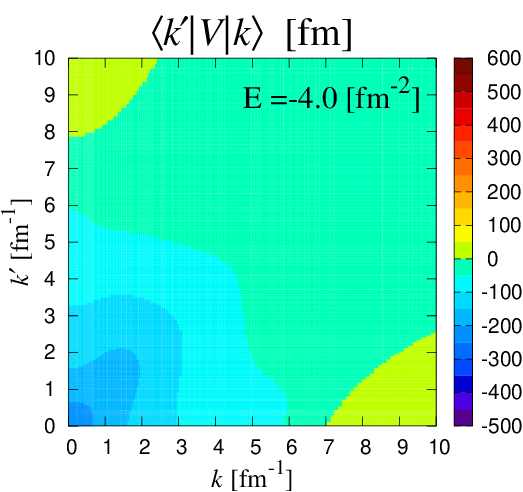}
\vspace{-5mm}
\caption{
SB-decomposition $V_{\rm S} + V_{\rm B} = V$ of $V$.
$\langle k' |V_{\rm S} | k  \rangle$ is 
calculated via Eq.(\ref{eqn:def_VPS}) using 
$\langle k' |T | k  \rangle$ in Fig.\ref{fig:delta_T_general}.
$\langle k' |V_{\rm B} | k  \rangle$ is taken 
from Fig.\ref{fig:inv_example_4} corresponding to $E= -4.0\, {\rm fm}^{-2}$.
Other notation is the same as for Fig.\ref{fig:inv_example_4}.
 See the text.}
\label{fig:general_example_4}
\end{figure}

\begin{figure}[!htb]
\vspace{10mm}
\threeDGraphs
{VS_changed_phase_v4.eps}
 {VPB_gamma2=1.eps}
{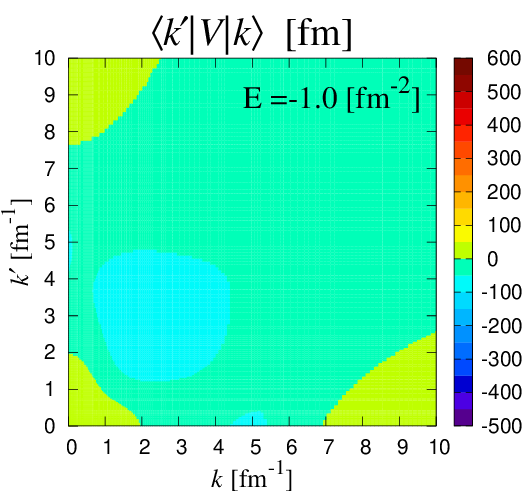}
\vspace{-5mm}
\caption{
$\langle k' |V_{\rm B} | k  \rangle$ is taken from Fig.\ref{fig:inv_example_1} 
corresponding to $E = -1.0\, {\rm fm}^{-2}$.
Other notation is the same as for Fig.\ref{fig:general_example_4}.
 See the text.}
\label{fig:general_example_1}
\end{figure}

\begin{figure}[!htb]
\vspace{10mm}
\threeDGraphs
{VS_changed_phase_v4.eps}
 {VPB_gamma2=0.eps}
{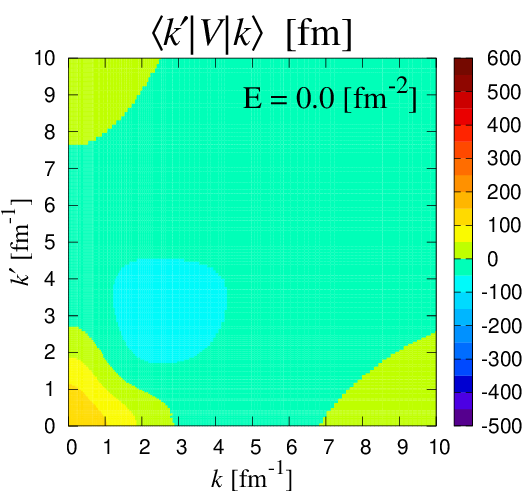}
\vspace{-5mm}
\caption{
$\langle k' |V_{\rm B} | k  \rangle$ is taken from Fig.\ref{fig:inv_example_0} 
corresponding to $E = 0.0\, {\rm fm}^{-2}$.
Other notation is the same as for Fig.\ref{fig:general_example_4}.
 See the text.}
\label{fig:general_example_0}
\end{figure}

\begin{figure}[!htb]
\vspace{10mm}
\threeDGraphs
{VS_changed_phase_v4.eps}
 {VPB_gamma2=-1.eps}
{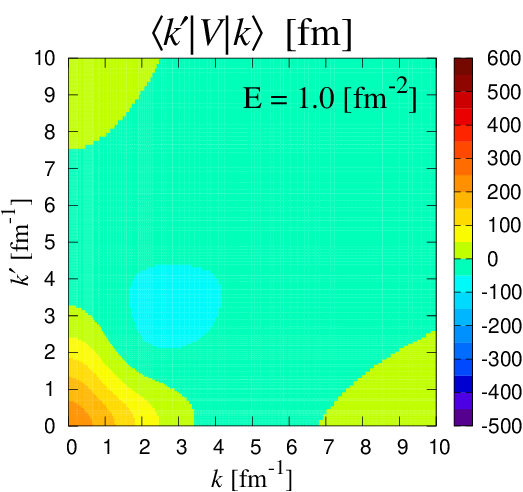}
\vspace{-5mm}
\caption{
$\langle k' |V_{\rm B} | k  \rangle$ is taken from Fig.\ref{fig:inv_example_minus_1}
corresponding to $E= 1.0\, {\rm fm}^{-2}$.
Other notation is the same as for Fig.\ref{fig:general_example_4}.
 See the text.}
\label{fig:general_example_minus_1}
\end{figure}

\begin{figure}[!htb]
\vspace{10mm}
\threeDGraphs
{VS_changed_phase_v4.eps}
 {VPB_gamma2=-4.eps}
{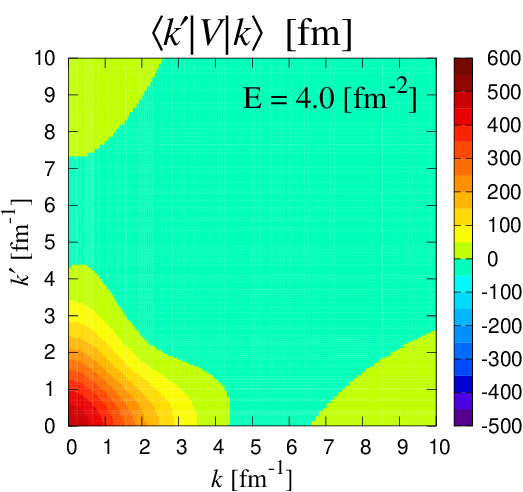}
\vspace{-5mm}
\caption{
$\langle k' |V_{\rm B} | k  \rangle$ is taken from Fig.\ref{fig:inv_example_minus_4}
corresponding to $E= 4.0\, {\rm fm}^{-2}$.
Other notation is the same as for Fig.\ref{fig:general_example_4}.
 See the text.}
\label{fig:general_example_minus_4}
\end{figure}

In step (ii), with $P_{\rm B} = |\phi \rangle \langle \phi |$, 
we solve the coupled set of Eqs.({\bf A}) and ({\bf B}) for $T$.
Because we find infinitely many solutions, we arbitrarily choose 
$\langle k' |T | k  \rangle$ in the left panel of Fig.\ref{fig:delta_T_general} 
which in turn gives the phase shift $\delta(k)$ in the right panel.
Note that they are different from those in Fig.\ref{fig:delta_T_Gauss}.
Therefore, the corresponding $\langle k' |V_{\rm S} | k  \rangle$
calculated  via Eq.(\ref{eqn:def_VPS}),
which is shown in the left panels of Figs.\ref{fig:general_example_4} -- \ref{fig:general_example_minus_4},  is different from $\langle k' |V_{0,{\rm S}} | k  \rangle$ 
in Figs.\ref{fig:inv_example_4} -- \ref{fig:inv_example_minus_4}.

In step (iii), we make $\langle k' |V_{\rm B} | k  \rangle$
in the same way as in Sect. \ref{sec:numerical_example_simple}; 
we consider five different eigenenergies of $| \phi \rangle$; 
$E= -4.0\, {\rm fm}^{-2}$,  
$-1.0\, {\rm fm}^{-2}$,
$0.0\, {\rm fm}^{-2}$,   
$1.0\, {\rm fm}^{-2}$, 
and  
$4.0\, {\rm fm}^{-2}$, 
each of which gives $\langle k' |V_{\rm B} | k  \rangle$  
in Figs.\ref{fig:inv_example_4} -- \ref{fig:inv_example_minus_4}
 in order.



Finally, by adding 
$\langle k' |V_{\rm S} | k  \rangle$ 
to  
each  $\langle k' |V_{\rm B} | k  \rangle$ 
in Figs.\ref{fig:inv_example_4} -- \ref{fig:inv_example_minus_4}, 
we obtain its corresponding 
$\langle k' |V_{\rm S} | k  \rangle + \langle k' |V_{\rm B} | k  \rangle = \langle k' |V | k  \rangle$
in Figs.\ref{fig:general_example_4} -- \ref{fig:general_example_minus_4}.

Let us note the following; 
 $\langle k' |V | k  \rangle$'s in Figs.\ref{fig:inv_example_minus_4}
and \ref{fig:general_example_minus_4}, for example, 
 share the same BIC $| \phi \rangle$ 
at the same energy ($E= 4.0\, {\rm fm}^{-2}$) 
because they share the same $\langle k' |V_{\rm B} | k  \rangle$.
On the other hand,  they give different $\langle k' |T | k  \rangle$'s
because their $\langle k' |V_{\rm S} | k  \rangle$'s are different. 
The same argument applies to each pair of
 Figs.\ref{fig:inv_example_4}
and \ref{fig:general_example_4}, 
Figs.\ref{fig:inv_example_1} and \ref{fig:general_example_1}, 
Figs.\ref{fig:inv_example_0} and \ref{fig:general_example_0},  and 
Figs.\ref{fig:inv_example_minus_1} and \ref{fig:general_example_minus_1}.
In other words, for each $\langle k' |T | k  \rangle$ which is compatible 
with the given bound state $| \phi \rangle$,\footnote{
This means that $\langle k' |T | k  \rangle$ 
is a solution to  the coupled set of Eqs.\,({\bf A}) and ({\bf B})
which are fixed  by  the given bound state $| \phi \rangle$ only,  
irrespective of its eigenenergy $E$ being negative or positive.
}
we can construct a Hermitian potential 
which supports $| \phi \rangle$
at any (positive or negative) eigenenergy as we like.
The above observation suggests 
that BIC's are as common as negative energy bound states
if we look into all Hermitian potentials.\footnote{
We shall soon make a rigorous proof of this statement 
in Sect. \ref{sec:generality_BIC}.
}

\section{Discussion on local potentials}
\label{sec:discussion_local}

In this and following sections, we make discussions.  
First,  in this section, we examine local potentials in connection with the BIC.
It is known  that a  local potential $V(r)$, which 
is continuous except at a finite number of finite discontinuities, 
gives the usual scattering theory if it  satisfies \cite{Newton82,Taylor83}\footnote{
The notation $V(r) = O(r^{p})$ means that $|V(r)| \leqq c \,|r|^p$ for some constant $c$.
} 
\begin{align}
{\rm I.} ~~&V(r) = O(r^{-3-\epsilon}), ~~\mbox{as}~ r \to \infty~~(\mbox{some}~\epsilon >0 ).
\label{eqn:condition_V_local_1}\\
{\rm II.} ~~&V(r) = O(r^{-2+\epsilon}), ~~\mbox{as}~ r \to 0~~(\mbox{some}~\epsilon >0 ).
\label{eqn:condition_V_local_2}
\end{align}

\noindent
In Sect.\,\ref{sec:rareness_local_BIC}, 
we shall see that if a  local potential $V(r)$ supports a BIC,
it  cannot satisfy condition I, 
and the BIC wave function $\phi(r)$ must have an oscillatory structure.
Then in Sect.\,\ref{sec:vonNeumann_Wigner_pot}, we show such an example; 
 we explain the only local potential we know which supports a BIC, i.e., 
the von Neumann and Wigner potential.

\subsection{Rarity of BIC for local potentials}
\label{sec:rareness_local_BIC}
Here we show the following two points.
\begin{itemize}
\item[(i)]
If a local potential $V(r)$ supports a BIC  $\phi(r)$, 
$V(r)$ does not satisfy condition I  of Eq.(\ref{eqn:condition_V_local_1}),
 and  $\phi(r)$ is an oscillating function with
an infinite number of zeros.   Accordingly, $V(r)$ also is an oscillating function.
\item[(ii)]
If a BIC wave function $\phi(r)$  has at most a finite number of zeros,
the potential $V$ of the system cannot be local but must be nonlocal, i.e., 
  $\langle r' | V  | r \rangle$ cannot be proportional to $\delta (r'-r)$.
\end{itemize}

Let us start with  point (i). 
First, we examine a case where a local potential $V(r)$ satisfying conditions I and II
 supports a BIC  
 wave function $\phi(r)$ at a positive energy $E=k^2$ via  
\begin{equation}
(-\nabla^2 +V(r))  \phi(r)  = k^2 \phi(r).
\label{eqn:Sch_V_local_1}
\end{equation}
Now we make an assumption that  $\phi(r)$ has only a finite number of zeros.
Then, because of condition I for $V(r)$, 
we can choose $R$ in such a way that   
 $\phi(r)$ has no zeros   for $r > R$ and  $V(r)  \phi(r)$ can be neglected there 
compared to $k^2 \phi(r)$ with any desired accuracy. 
This means that the solution $\phi(r)$ of Eq.(\ref{eqn:Sch_V_local_1}) for $r > R$ 
can be approximated by $j_0(kr+\alpha)$, with $\alpha$ being an appropriate phase.
However, because $j_0(kr+\alpha)$ has an infinite number of zeros 
in the range  $r > R$, the above solution $j_0(kr+\alpha)$  contradicts 
our choice of $R$.    By reduction to absurdity, therefore,  
we can conclude that  our assumption is false, i.e., 
 $\phi(r)$ is an oscillating function 
with an infinite number of zeros, leading to the second half of  point (i). 

Second, taking into account the above observation, 
let us express   $\phi(r)$ of Eq.(\ref{eqn:Sch_V_local_1}) as
$\phi(r) = j_{0}(kr+\alpha) f(r)$ with $f(r)$ being an unknown envelope function
which guarantees the normalizability of $\phi(r)$.   Then, it is easy to see\footnote{
Note that $\nabla^2 = \dfrac{{\rm d}^2}{{\rm d}r^2} + \dfrac{2}{r}\dfrac{{\rm d}}{{\rm d}r}$.
}
\begin{equation}
-\nabla^2  \phi(r)  = k^2 \phi(r) 
- 2 \frac{{\rm d} j_0(kr+\alpha)}{{\rm d} r} \frac{{\rm d} f(r)}{{\rm d} r}
- j_0(kr+\alpha) \frac{{\rm d}^2 f(r)}{{\rm d} r^2}
- j_0(kr+\alpha) \frac{2}{r}\frac{{\rm d} f(r)}{{\rm d}r }.
\label{eqn:Sch_V_local_2}
\end{equation}
Now we suppose that  $f(r) = O(r^{-n})$  with some $n\geqq 1$ for  $r \to \infty$.\footnote{
It can be easily shown \cite{Stillinger_PRA_75} that an exponentially decaying $f(r)$ is not
consistent with  Schr\"odinger  equation (\ref{eqn:Sch_V_local_1}).
}
This means that the leading contribution 
on the right hand side as  $r \to \infty$ comes from the second term; 
\begin{equation}
- 2 \frac{{\rm d} j_0(kr+\alpha)}{{\rm d} r} \frac{{\rm d} f(r)}{{\rm d} r} ~=~ O(r^{-n-1}), 
~~\mbox{as}~r \to \infty.
\label{eqn:Sch_V_local_3}
\end{equation}
Then, by comparing Eqs.(\ref{eqn:Sch_V_local_1}) and (\ref{eqn:Sch_V_local_2}),
we realize that $V(r) = O(r^{-1})$ as  $r \to \infty$.
The above observation shows the first half, 
to complete a proof of point (i).

It is now easy to show point (ii) as a contraposition to point (i).
We can thus prove both points (i) and (ii). 
Note that point (i) suggests that 
a BIC can rarely occur in a local potential.\footnote{
This point shall be proven later in a definite manner.}

\subsection{Existing theory -- von Neumann and Wigner potential}
\label{sec:vonNeumann_Wigner_pot}
 In Ref.\,\cite{vonNeumann29}, von Neumann and Wigner 
first showed that a BIC is possible by 
deriving a local potential $V(r)$ which conforms to point (i)
in Sect.\,\ref{sec:rareness_local_BIC}.
As briefly explained in Appendix \ref{app:vonNeumann_Wigner_pot}, 
their potential  $V(r)$  in the S-wave channel describes  
$\phi(r) = j_{0}(kr) f(r)$, the spherical Bessel function $j_{0}(kr)$ in an envelope $f(r)$,
as a BIC at a  positive energy $E=k^2$.
Comparing our theory to theirs, we realize the followings.
\begin{itemize}
\item[$\bullet$]
The wave function of their  BIC  is tailored 
from the oscillating plane wave state $j_{0}(kr)$.  
 Correspondingly, their potential also has an oscillatory structure, 
\begin{equation}
 V(r) \propto \frac{\sin 2kr}{2kr},~~~\mbox{as}~r \to \infty, 
\label{eqn:pot_vNW_asymptotic}
\end{equation}
which does not satisfy condition I  of Eq.(\ref{eqn:condition_V_local_1}) 
in accordance with point (i) in Sect.\,\ref{sec:rareness_local_BIC}, 
leading to a destructive interference which allows for the existence of the BIC. 
In our theory, 
on the other hand, we can choose any normalizable wave function 
for the BIC.
Correspondingly, our BIC-supporting potential is not necessarily 
of an oscillatory structure, which can be confirmed clearly by 
Eq.(\ref{eqn:transformed_gauss_r}).

\item[$\bullet$]
The eigenenergy of their BIC,  $\phi(r) = j_{0}(kr) f(r)$, is fixed at $E=k^2$.
In our theory, on the other hand, we can  assign any eigenenergy $E$ to our 
 bound state.

\item[$\bullet$]
Except for the BIC  $\phi(r) = j_{0}(kr) f(r)$ at $E=k^2$, 
their theory does not give any information
 on scattering states and  other possible bound states.\footnote{
Note that, because their potential does not satisfy condition I 
of Eq.(\ref{eqn:condition_V_local_1}), we cannot even expect the usual 
description of scattering states. 
}
On the other hand, as explained in Sect.\,\ref{sec:general_design_V}, 
our theory keeps all eigenstates of the system under control by construction.
\end{itemize}

From the above observation, 
we realize that  the local potential of von Neumann and Wigner
is a very special case which cannot be generalized to 
other BIC's.
Further, because their BIC  is vulnerable to
any local perturbation \cite{Pursey_95, Weber_98},
we conclude that their BIC is an exceptional phenomenon
 in the group of all local potentials, which might accord with our intuition.
In the following, however, we shall see that the present theory predicts the opposite
if we survey the group of all  Hermitian potentials.

\section{Discussion on general nonlocal potentials}
\label{sec:discussion}
In this section, we make discussions on various aspects of the present theory
which treats general nonlocal potentials.

\subsection{Generality of BIC}
\label{sec:generality_BIC}

 In Sect.\,\ref{sec:transform_VB}, 
 we examined the transformation of $\langle k' |V_0 | k  \rangle 
= \langle k' |V_{0,{\rm S}} | k  \rangle + \langle k' |V_{0,{\rm B}} | k  \rangle$ 
into 
$\langle k' |V | k  \rangle 
= \langle k' |V_{0,{\rm S}} | k  \rangle + \langle k' |V_{\rm B} | k  \rangle$,  
which is controlled by the single variable $E$ 
assigned to  the normalizable state $\phi(k)$.
Depending on  $E$ being negative or positive,
$\langle k' |V | k  \rangle$ describes  
$\phi(k)$ as a negative or positive energy bound state.  
Therefore, starting from 
$\langle k' |V_0 | k  \rangle 
= \langle k' |V_{0,{\rm S}} | k  \rangle + \langle k' |V_{0,{\rm B}} | k  \rangle$  at hand, 
we can make as many potentials which present  $\phi(k)$ as a BIC as
those which describe  $\phi(k)$ as a negative energy bound state. 

To be precise, 
let $\mathbb{V}(T_0,\phi)$ be a group of all Hermitian potentials that 
share the half-on-shell T-matrix $\langle k' |T_0 | k  \rangle$ and
the single bound state  $| \phi  \rangle$.
We distinguish each $V \in  \mathbb{V}(T_0,\phi)$
 by the eigenenergy $E$ of  $| \phi  \rangle$.\footnote{
In the group $\mathbb{V}(T_0,\phi)$, 
$ \langle k' |V_{\rm S} | k  \rangle =  \langle k' |V_{0,{\rm S}} | k  \rangle$ is fixed,
and $\langle k' |V_{\rm B} | k  \rangle$ of Eq.(\ref{eqn:def_VPB_single_E}) alone varies
with $E$. 
Note that $V_0$ in Eq.(\ref{eqn:gauss_k}) is an element of $\mathbb{V}(T_0,\phi)$
specified by $E=-\gamma_0^2=-5.373\, {\rm fm}^{-2}$.
} 
Then, the above argument shows that half  elements of $\mathbb{V}(T_0,\phi)$ support
 $| \phi  \rangle$ at positive energies, and the other half at negative energies.  
Now, let $\mathbb{V}$ represent the group of all Hermitian potentials 
which have a single bound state. 
Obviously,  $\mathbb{V}$ is a direct sum of all possible subgroups defined similarly to  
$\mathbb{V}(T_0,\phi)$
in terms of a half-on-shell T-matrix and a bound state wave function, i.e.,\footnote{
In Eq.(\ref{eqn:V_direct_sum}), 
the half-on-shell T-matrices $\{T_0, T_1, T_2, \cdots \}$ are ``compatible'' with the bound state 
$| \phi \rangle$, i.e.,  they are solutions to  the coupled set of Eqs.\,({\bf A}) and ({\bf B})
for  the given bound state $| \phi \rangle$,  
 and $\{T'_0, T'_1, T'_2, \cdots \}$ are ``compatible'' with another bound state $| \phi' \rangle$, and so on.
}
\begin{align}
\mathbb{V} 
&= ~~~\mathbb{V}(T_0,\phi) \oplus \mathbb{V}(T_1,\phi) \oplus \mathbb{V}(T_2,\phi)\oplus \cdots
\nonumber\\
&~~~\oplus \mathbb{V}(T'_0,\phi') \oplus \mathbb{V}(T'_1,\phi') \oplus \mathbb{V}(T'_2,\phi')\oplus \cdots
\nonumber\\
&~~~\oplus \mathbb{V}(T''_0,\phi'') \oplus \mathbb{V}(T''_1,\phi'') \oplus \mathbb{V}(T''_2,\phi'')\oplus \cdots.
\label{eqn:V_direct_sum}
\end{align}
Exactly in the same way as $\mathbb{V}(T_0,\phi)$, 
each subgroup on the right hand side of Eq.(\ref{eqn:V_direct_sum}) 
contains as many potentials with a BIC
as those without a BIC, so does $\mathbb{V}$ on the left hand side.
The above argument can be easily generalized to 
systems with two or more bound states,
which immediately leads us to the following conclusion;
if we examine all Hermitian potentials, most of which are  nonlocal,  
we would find  as many  BIC's as negative energy bound states.    
Surprisingly enough,  in the group of all possible Hermitian potentials, 
the BIC is not at all an exceptional phenomenon as predicted 
by local potentials  in Sect.\,\ref{sec:discussion_local}.

\subsection{Coexistence of BIC and scattering state}
\label{sec:existence_BIC}

Using the method in Sect.\,\ref{sec:simple_form_V_supporting_BIC},
let us prepare  
a potential    $\langle k' |V| k  \rangle$ 
which supports a BIC $| \phi \rangle$ at $E=K^2$.
Then, 
 the scattering state $| K \rangle\!_+ $ 
and the BIC $| \phi \rangle$ are degenerate at $E=K^2$ in our system.\footnote{
The original potential $V_0$ describes the scattering state $| K \rangle\!_+ $  at $E=K^2$  
and the bound state $| \phi \rangle$ at $E= -\gamma_0^2$,
which are of course mutually orthogonal.
The transformed potential $V$ describes the scattering state $| K \rangle\!_+ $
exactly in the same way as $V_0$ by construction.   
At the same time, it supports  $| \phi \rangle$ at $E= K^2$
as explained in Sect.\,\ref{sec:simple_form_V_supporting_BIC}.
Therefore, $| K \rangle\!_+ $ and $| \phi \rangle$ 
 are eigenstates of $H_0+V$ with the same eigenenergy, $E=K^2$.
} 
This means that the Schr\"odinger equation  
in coordinate space
corresponding to Eq.(\ref{eqn:Sch_BIC}), 
which is the following integro-differential equation for the S-wave channel,\footnote{
$ \langle r |V| r'  \rangle= V(r,r')$ is generally nonlocal.    See Eq.(\ref{eqn:transformed_gauss_r}).
}
\begin{equation}
\left(-\dfrac{{\rm d}^2}{{\rm d}r^2} - \dfrac{2}{r}\dfrac{{\rm d}}{{\rm d}r} \right)   \phi(r)
+ \int_0^{\infty} \! r'^2{\rm d}r'  \, V(r,r') \phi (r') = K^2 \phi(r),
\label{eqn:Sch_general_coordinate}
\end{equation}
has two different solutions  which are regular at the origin;
one is the BIC  $\phi(r)$, and the other 
 is the scattering state $\psi_{K}^{(+)}(r) = \langle r| K \rangle\!_+$
with the outgoing boundary condition.
This degeneracy implies the existence of 
a nontrivial solution $\chi(r)$ to Eq.(\ref{eqn:Sch_general_coordinate})
 that satisfies $\chi(0) = \chi'(0)=0$ where $\chi'(r) = \dfrac{{\rm d}}{{\rm d}r}\chi(r)$
\cite{Mulligan_76, Mulligan_81, Warke_71}.\footnote{
In fact, 
we can prepare such a solution $\chi(r)$ as an appropriate linear combination of 
$\phi(r)$ of the BIC and $\psi_{K}^{(+)}(r)$
for our BIC-supporting potential.
}
Therefore, the conclusion in Sect.\,\ref{sec:generality_BIC} means that,  
for an arbitrary $\phi(r)$
at an arbitrary positive energy   $E=K^2$,  there is 
a  nonlocal potential $V(r,r')$  which allows a solution  
 $\phi(r) = \chi(r)$ to Eq.(\ref{eqn:Sch_general_coordinate})
 satisfying $\chi(0) = \chi'(0)=0$.

In  a system described by a local potential $V(r)$, however, 
the situation is different.
Here, the Schr\"odinger equation  in coordinate space 
is given by the second order differential equation:
\begin{equation}
\left(-\dfrac{{\rm d}^2}{{\rm d}r^2} - \dfrac{2}{r}\dfrac{{\rm d}}{{\rm d}r} \right)   \phi(r)
+ V(r) \phi (r) = K^2 \phi(r).
\label{eqn:Sch_general_coordinate_local}
\end{equation}
It is straightforward to show that, using an argument of the uniqueness of solution
for ordinary differential equation,  
 Eq.(\ref{eqn:Sch_general_coordinate_local}) 
with the initial condition $\phi(0) = \phi'(0)=0$
 would generally yield the unique solution  $\phi(r) \equiv 0$.
In other words, 
a local potential $V(r)$ would not generally allow the degeneracy
of the scattering state $\psi_{K}^{(+)}(r) = \langle r| K \rangle\!_+$ and 
the BIC $\phi(r)$ at  $E=K^2$,
which in turn means that the BIC is a rare phenomenon for local potentials.

Summarizing the above,
 nonlocal potentials easily allow the coexistence of 
the scattering state $\psi_{K}^{(+)}(r)$ and the BIC $\phi(r)$ at the same energy 
$E=K^2$, while local potentials rarely do. 
Consequently,  BIC is a commonplace event in the group of all Hermitian potentials 
most of which are nonlocal,
while it is not the case in its subgroup composed of local potentials only.

\section{Local versus nonlocal potentials}
\label{sec:local_vs_nonlocal}

In Sect.\,\ref{sec:existence_BIC}, we have seen the 
difference of the Schr\"odinger equation between local and nonlocal potentials 
in connection with the BIC.  
Here we compare local and nonlocal potentials in terms of the
SB-decomposition of Sect.\,\ref{sec:general_design_V},
which will show the importance of nonlocal potentials in the study of BIC
from another point of view.

\subsection{Potential group $\mathbb{V}(t,\phi)$}
\label{sec:local_vs_nonlocal_0}

Let us specify a group of potentials we are going to examine.
\begin{itemize}
\item[$\bullet$]
 We choose  an on-shell T-matrix 
\begin{equation}
t(k)=-\dfrac{1}{\pi \rho_k} {\rm e}^{i\delta(k)} \sin \delta(k) ,
\label{eqn:tk_via_phase_shift}
\end{equation}
by  arbitrarily giving a phase shift $\delta(k)$ which is
in accordance with Levinson's theorem.\footnote{
$\rho_k = \dfrac{k}{2}\dfrac{1}{(2\pi)^3}$ is the density of states.
}
\item[$\bullet$]
 We fix  a single bound state $|\phi \rangle$ in such a way 
that it behaves as\footnote{
This choice of $|\phi \rangle$ excludes the 
 von Neumann and Wigner potential in  Sect.\,\ref{sec:vonNeumann_Wigner_pot}.
} 
\begin{equation}
\phi(r)  \propto \exp(-\gamma r),~~ \mbox{as}~ r \to \infty.
\label{eqn:phi_asymptotic}
\end{equation}
\end{itemize}
Now we define a group $\mathbb{V}(t,\phi)$; 
 it is a group of all Hermitian potentials that 
share the on-shell T-matrix $t(k)$ of Eq.(\ref{eqn:tk_via_phase_shift}) and
the single bound state  $| \phi  \rangle$ satisfying Eq.(\ref{eqn:phi_asymptotic}).
Note that $\mathbb{V}(t,\phi)$ is the group 
of all solutions to the inverse scattering problem 
for $t(k)$ with the bound state $| \phi  \rangle$.
In the following,  we  examine $\mathbb{V}(t,\phi)$ 
in terms of the SB-decomposition of Sect.\,\ref{sec:general_design_V}.

First, for any value of the eigenenergy $E$ of $| \phi  \rangle$,
Eq.(\ref{eqn:def_VPB_single_E}) gives its corresponding $V_{\rm B}(\phi,E)$.\footnote{
The present notation $V_{\rm B}(\phi,E)$ explicitly shows 
that $V_{\rm B}$ is determined by $\phi$ and $E$.
}
We define, therefore,  
\begin{equation}
\mathbb{V}_{\rm B}(\phi)=\{V_{\rm B}(\phi,E)\}, 
\label{eqn:def_group_VB_phi}
\end{equation}
as  a  group composed of $V_{\rm B}(\phi,E)$'s. 
Note that, because $E$ is a continuous parameter, 
$\mathbb{V}_{\rm B}(\phi)$
contains infinite number of elements.

Second, for the fixed $P_{\rm B}=|\phi \rangle  \langle \phi  |$, let us take a
solution $T$ of  Eqs.\,({\bf A}) and ({\bf B})
with the boundary condition $\langle k | T  | k \rangle$ being equal to $t(k)$
of  Eq.(\ref{eqn:tk_via_phase_shift}),
which in turn gives  its corresponding $V_{\rm S}(T,\phi)$
via Eq.(\ref{eqn:def_VPS}).\footnote{
The notation $V_{\rm S}(T,\phi)$ 
shows explicitly that $V_{\rm S}$ is determined by $T$ and $\phi$.
}
Then we define 
\begin{equation}
\mathbb{V}_{\rm S}(t,\phi)=\{V_{\rm S}(T,\phi) \}, 
\label{eqn:def_group_VS_t_phi}
\end{equation}
as a group of all possible $V_{\rm S}(T,\phi)$'s thus obtained.\footnote{
Different elements of $\mathbb{V}_{\rm S}(t,\phi)$ are distinguished by different $T$'s
which share the on-shell T-matrix $\langle k | T  | k \rangle=t(k)$.
By writing such $T$'s as $T_1,T_2,\cdots$, we can express the group as 
$\mathbb{V}_{\rm S}(t,\phi)=\{V_{\rm S}(T_1,\phi),  V_{\rm S}(T_2,\phi), \cdots\}$.  
}
Since  an  infinite  number of $T$'s would satisfy  Eqs.\,({\bf A}) and ({\bf B})
with the boundary condition $\langle k | T  | k \rangle=t(k)$,
$\mathbb{V}_{\rm S}(t,\phi)$ 
would contain infinitely many elements \cite{Takayanagi_PTEP_23}.\footnote{
In the space orthogonal to $|\phi \rangle$,  
 there are  infinite number of different complete orthonormal sets 
which share the same asymptotic behaviors (or $t(k)$).  
Therefore, 
because knowledge of $T$ is equivalent to that of a
complete orthonormal set $\{| k \rangle\!_+ \}$ of scattering states,  
 we can find infinitely many $T$'s satisfying Eqs.\,({\bf A}) and ({\bf B}).
}

\begin{figure}[!htb]
\vspace{2mm}
\hspace{10mm}\includegraphics[width=0.85\columnwidth]{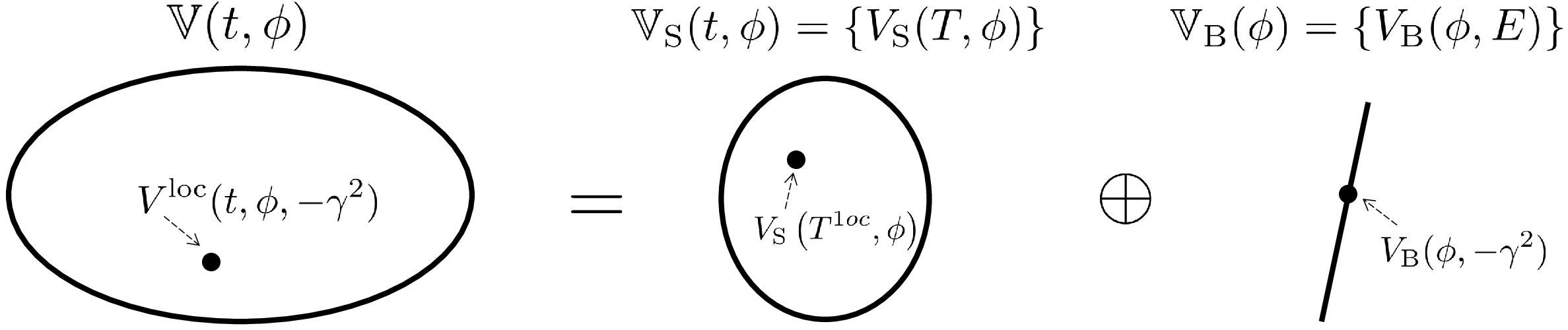}
\caption{
Schematic expression for Eq.(\ref{eqn:V_direct_sum_2}).
$\mathbb{V}_{\rm B}(\phi)$ is denoted by a single line
showing that it is a group with a single parameter $E$.
The only local potential in $\mathbb{V}(t,\phi)$,
given by  Eq.(\ref{eqn:V_local_direct_sum}), 
is  indicated in the figure. 
See the text.}
\label{fig:direct_sum_1}
\end{figure}	

Finally,  the SB-decomposition expresses each 
 potential in $\mathbb{V}(t,\phi)$
uniquely as
\begin{equation}
V(T,\phi,E) =V_{\rm S}(T,\phi) + V_{\rm B}(\phi,E).
\label{eqn:V_direct_sum_3}
\end{equation}
Correspondingly,  $\mathbb{V}(t,\phi)$  is  
a direct sum of $\mathbb{V}_{\rm S}(t,\phi)$ and $\mathbb{V}_{\rm B}(\phi)$, i.e., 
\begin{equation}
\mathbb{V}(t,\phi) = \mathbb{V}_{\rm S}(t,\phi) \oplus \mathbb{V}_{\rm B}(\phi),
\label{eqn:V_direct_sum_2}
\end{equation}
as shown in Fig.\,\ref{fig:direct_sum_1}.\footnote{
Note that $\mathbb{V}(t,\phi)$ of Eq.(\ref{eqn:V_direct_sum_2})
is much larger than $\mathbb{V}(T,\phi)$ introduced in Sect.\,\ref{sec:generality_BIC}; 
by choosing a single $V_{\rm S}(T,\phi) \in \mathbb{V}_{\rm S}(t,\phi)$
in Eq.(\ref{eqn:V_direct_sum_2}), the group $\mathbb{V}(t,\phi)$
reduces to $\mathbb{V}(T,\phi)$.
}
In the following, we look for local and nonlocal potentials in  
$\mathbb{V}(t,\phi)$ of Eq.(\ref{eqn:V_direct_sum_2}).

\subsection{Local  potentials}
\label{sec:local_vs_nonlocal_1}

Here, we study a possible local potential in $\mathbb{V}(t,\phi)$.
In the group of all local potentials, one can rarely find a BIC-supporting potential
as explained from the viewpoint of 
the differential equation in Sect.\,\ref{sec:existence_BIC}.\footnote{
In fact, we know only a single BIC-supporting local potential,  
the von Neumann and Wigner potential in  Sect.\,\ref{sec:vonNeumann_Wigner_pot} 
of which the bound state does not satisfy 
the assumption  of an exponentially damping wave function in Eq.(\ref{eqn:phi_asymptotic}).
}
Here, we explain the above situation from another point of view;
we shall see that there is no BIC-supporting local potential  in $\mathbb{V}(t,\phi)$.

First, we start with the group $\mathbb{V}_{\rm B}(\phi)$ of Eq.(\ref{eqn:def_group_VB_phi}).
For a local potential satisfying condition I of Eq.(\ref{eqn:condition_V_local_1}),
Eq.(\ref{eqn:phi_asymptotic}) means  that 
$\phi(r)$ has the eigenenergy $E=-\gamma^2$, and therefore
that $\phi(r)$ is not  a BIC.  
Then,  Eq.(\ref{eqn:def_VPB_single_E}) uniquely picks up 
an element $V_{\rm B}(\phi,-\gamma^2) \in \mathbb{V}_{\rm B}(\phi)$.

Second, we turn to the group $\mathbb{V}_{\rm S}(t,\phi)$ 
of Eq.(\ref{eqn:def_group_VS_t_phi}).
Let us recall the well-known result
 of the inverse scattering problem for local potentials \cite{Newton82,Chadan89};
there is only a single  local solution $V^{\rm loc}(t,\phi, -\gamma^2; r)$
 to the inverse scattering problem  for a given set of $\{t,\phi, -\gamma^2\}$.\footnote{
To be precise, 
existence of the solution requires the phase shift to be ``reasonable'' \cite{Newton82}.
}
Therefore, the only local potential $V^{\rm loc}(t,\phi, -\gamma^2)$ 
uniquely chooses its corresponding element 
 $V_{\rm S}(T^{\rm loc},\phi) \in \mathbb{V}_{\rm S}(t,\phi)$,\footnote{
$T^{\rm loc}$ is the half-on-shell T-matrix corresponding to $V^{\rm loc}(t,\phi, -\gamma^2)$.
}
 to compose  $V^{\rm loc}(t,\phi, -\gamma^2)$ in the form of Eq.(\ref{eqn:V_direct_sum_3}) as 
\begin{equation}
V^{\rm loc}(t,\phi, -\gamma^2) 
= V_{\rm S}(T^{\rm loc},\phi) + V_{\rm B}(\phi,-\gamma^2).
\label{eqn:V_local_direct_sum}
\end{equation}

\noindent
For the sake of clarity, we have embedded  Eq.(\ref{eqn:V_local_direct_sum})  
in Fig.\,\ref{fig:direct_sum_1} for Eq.(\ref{eqn:V_direct_sum_2}).
Each element  $V(T,\phi,E) \in \mathbb{V}(t,\phi)$
corresponds to a specific choice of $V_{\rm S}(T,\phi) \in \mathbb{V}_{\rm S}(t,\phi)$
and $V_{\rm B}(\phi,E) \in \mathbb{V}_{\rm B}(\phi)$,
where both $\mathbb{V}_{\rm S}(t,\phi)$ and $\mathbb{V}_{\rm B}(\phi)$
contain infinite number of elements.
Among all these combinations in Fig.\,\ref{fig:direct_sum_1},  only a single choice,   
  $V_{\rm S}(T^{\rm loc},\phi) \in \mathbb{V}_{\rm S}(t,\phi)$
and $V_{\rm B}(\phi,-\gamma^2) \in \mathbb{V}_{\rm B}(\phi)$, 
makes the only local potential in $\mathbb{V}(t,\phi)$.

To summarize, with condition I of Eq.(\ref{eqn:condition_V_local_1}) for the potential and 
the assumption  of an exponentially damping wave function in Eq.(\ref{eqn:phi_asymptotic}), 
 $V^{\rm loc}(t,\phi, -\gamma^2)$ of Eq.(\ref{eqn:V_local_direct_sum})
is the only local potential in  $\mathbb{V}(t,\phi)$,
and its bound state $| \phi  \rangle$ is not a BIC.

\subsection{Nonlocal  potentials}
\label{sec:local_vs_nonlocal_2}

Here, we turn to nonlocal potentials in $\mathbb{V}(t,\phi)$.
Note that 
all elements of  $\mathbb{V}(t,\phi)$ but $V^{\rm loc}(t,\phi, -\gamma^2)$ 
of Eq.(\ref{eqn:V_local_direct_sum}) are nonlocal potentials.

Let us arbitrarily choose two elements,    
 $V_{\rm S}(T,\phi) \in \mathbb{V}_{\rm S}(t,\phi)$ and 
$V_{\rm B}(\phi,E) \in \mathbb{V}_{\rm B}(\phi)$,
to compose $V(T,\phi,E) \in \mathbb{V}(t,\phi)$ of Eq.(\ref{eqn:V_direct_sum_3}).
Then, depending on $E<0$ or $E>0$, our potential $V(T,\phi,E)$ describes 
$| \phi  \rangle$ as a negative or positive energy bound state, 
irrespective of the choice of  $V_{\rm S}(T,\phi) \in \mathbb{V}_{\rm S}(t,\phi)$.
We can conclude, therefore,  
that half elements of $\mathbb{V}(t,\phi)$ 
carry positive $E$ and   describe $| \phi  \rangle$ as a BIC,\footnote{
This is another way to confirm the results of Sect.\,\ref{sec:generality_BIC}, 
the generality of BIC.  
}
 and they are all nonlocal potentials.

\subsection{Importance of nonlocal potential}
\label{sec:local_vs_nonlocal_3}

Let $\mathbb{V}$ represent 
the group of all Hermitian potentials with a single bound state.\footnote{
We assume condition I of Eq.(\ref{eqn:condition_V_local_1}) and 
an exponentially damping wave function of the bound state as in Eq.(\ref{eqn:phi_asymptotic}).
} 
Then, $\mathbb{V}$ is a direct sum of all subgroups
defined similarly to $\mathbb{V}(t,\phi)$ of Eq.(\ref{eqn:V_direct_sum_2}), i.e.,
\begin{align}
\mathbb{V} 
&= ~~~\mathbb{V}(t_0,\phi) \oplus \mathbb{V}(t_1,\phi) \oplus \mathbb{V}(t_2,\phi)\oplus \cdots
\nonumber\\
&~~~\oplus \mathbb{V}(t_0,\phi') \oplus \mathbb{V}(t_1,\phi') \oplus \mathbb{V}(t_2,\phi')\oplus \cdots
\nonumber\\
&~~~\oplus \mathbb{V}(t_0,\phi'') \oplus \mathbb{V}(t_1,\phi'') \oplus \mathbb{V}(t_2,\phi'')\oplus \cdots.
\label{eqn:V_local_direct_sum_2}
\end{align}

\noindent
In Sect.\,\ref{sec:local_vs_nonlocal_1},
we have seen  that there is at most a single local potential in each subgroup
on the right hand side of Eq.(\ref{eqn:V_local_direct_sum_2}),\footnote{
Potentials in $\mathbb{V}(t_0,\phi)$ and $\mathbb{V}(t_0,\phi')$ carry the same 
on-shell T-matrix $t_0$.    However, they cannot share the same half-on-shell T-matrix;
each potential in $\mathbb{V}(t_0,\phi)$ requires 
a half-on-shell T-matrix satisfying Eqs.({\bf A}) and ({\bf B}) 
with $P_{\rm B}=|\phi \rangle \langle \phi |$,
while  $\mathbb{V}(t_0,\phi')$ corresponds to
 different $P_{\rm B}=|\phi' \rangle \langle \phi' |$.
}
 and it does not  support a BIC. 
On the other hand, 
in Sect.\,\ref{sec:local_vs_nonlocal_2}, we have shown that 
half elements of each subgroup, which are  all nonlocal potentials,
 support  a BIC.
In the search for BIC, therefore, 
we have to examine nonlocal potentials.

\section{Observing BIC}
\label{sec:observing_BIC}

Here we discuss how we can observe a BIC or confirm its existence
in a system described by a nonlocal potential.    

To begin with, let us  note the difficulty to observe a BIC directly.
On the experimental side,  we can observe a BIC only via its interaction with some other systems.
On the other hand, however, the interaction necessarily gives a finite width to the BIC,
which would inevitably obscure 
the BIC to make its empirical identification difficult.
On the theoretical side,  we note that  knowledge of $\langle k' | T  | k  \rangle $ alone 
does not tell any information on a possible BIC of the system.
We can easily prepare two  different systems with the same $\langle k' | T  | k  \rangle $, 
one of which supports a bound state  $| \phi \rangle$ as a BIC at a positive energy $E=K^2$,
and the other at a negative energy $E=-\gamma^2$.\footnote{
An example of the pair of such systems can be found in 
Sect.\,\ref{sec:results_VS_plus_VB_makes_V}; 
potentials in Figs.\ref{fig:inv_example_4} and \ref{fig:inv_example_minus_4}
 yield the same $\langle k' | T  | k  \rangle$
but give the common bound state $| \phi \rangle$ at different energies
 $E=-4.0\, {\rm fm}^{-2}$ and $4.0\, {\rm fm}^{-2}$.
In the SB-decomposition, 
these potentials   
 have the same $V_{\rm S}$ but differ from each other in $V_{\rm B}$.
Generally, such a pair of systems can be prepared easily 
by the method in Sect.\,\ref{sec:simple_form_V_supporting_BIC}.
}
Naturally, we cannot distinguish between these two systems 
from their common $\langle k' | T  | k  \rangle$ only.\footnote{
Their common $T$ determines their common $V_{\rm S}$ only, 
leaving $V_{\rm B}$, and therefore $V= V_{\rm S} + V_{\rm B}$ of the system,  undetermined.
}
Therefore,  if we look into  the scattering data only, 
we cannot recognize the BIC or its precursor theoretically.

This is not, however, the whole story, as we explain below.

\subsection{Finding $N_{+}$}
\label{sec:observing_BIC_1}

It is known that Levinson's theorem, $\delta(0) -\delta(\infty)  = N \pi$,
holds not only for local but also for nonlocal potentials \cite{Chadan89,Martin_58},
where $N$ is the number of normalizable eigenstates of the system.
Let us express $N$ as  $N=N_{-} + N_{+}$ 
where $N_{-}$ is the number  of negative energy bound states
and $N_{+}$ is the number of BIC's.
Then,  as we show below, $N$ and $N_{-}$ are accessible both theoretically and experimentally,
and so is $N_{+}= N-N_{-}$.

First, we explain the experimental observation of $N_{+}$ 
using an example.
Suppose that, for a given system,     
 we have observed experimentally the phase shift $\delta(k)$ 
 in Fig.\ref{fig:delta_T_Gauss} which shows $N=1$,
and also that we do not find a negative energy bound state, i.e., $N_{-}=0$.
Then, we can  confirm  experimentally that $N_{+}= N-N_{-}=1$, 
and therefore 
 conclude  that the system supports a single  BIC.

Second, let us turn to the theoretical observation of $N_{+}$.
Here, the problem is how to determine $N_{+}$ of a system
of which the potential is known.
To be concrete, we  consider a system with $\langle k' |V | k  \rangle$
in Fig.\,\ref{fig:inv_example_minus_4}.
Then, we can easily calculate  its phase shift $\delta(k)$ in Fig.\,\ref{fig:delta_T_Gauss},
 which in turn gives $N=1$ via Levinson's theorem.    
At the same time,  we can confirm  $N_{-}=0$ by diagonalizing 
the Hamiltonian of Eq.(\ref{eqn:def_H}) in momentum space.    
We can thus obtain $N_{+}= N-N_{-}=1$ theoretically, and conclude 
that the system under consideration supports a single  BIC.\footnote{
When $N=1$, we can easily distinguish $N_{+}=1$ and $N_{-}=1$ cases
 by looking into the sign of $\langle 0 |V_{\rm B} | 0  \rangle$ only.
For a given $\langle k' |V | k  \rangle$, we can easily 
calculate  $\langle k' |T | k  \rangle$ and therefore $\langle k' |V_{\rm S} | k  \rangle$,
which in turn gives
 $\langle k' |V_{\rm B} | k  \rangle  = \langle k' |V | k  \rangle- \langle k' |V_{\rm S} | k  \rangle$.
Then, noting that 
$\langle 0 |V_{\rm B} | 0  \rangle = E \vert \langle 0 | \phi \rangle \vert^2$,
which is shown by Eq.(\ref{eqn:def_VPB_single_E}), 
has the same sign as the eigenenergy $E$,
we realize that $\langle 0 |V_{\rm B} | 0  \rangle > 0 $ means $N_{+}=1$,
and that $\langle 0 |V_{\rm B} | 0  \rangle < 0 $ implies $N_{-}=1$.
The above criterion can be easily confirmed, e.g., 
in Figs.\,\ref{fig:general_example_4}, \ref{fig:general_example_1},
\ref{fig:general_example_minus_1}, and \ref{fig:general_example_minus_4}.
}

 To summarize the above, by making use not only of the scattering data   $\delta(k)$,
but also of  the number $N_{-}$ of negative energy bound states which we can 
observe easily both experimentally and theoretically, 
we can obtain,
by virtue of Levinson's theorem, the number $N_{+}$ of BIC's in the system.

\subsection{Finding $\phi(k)$ and $K^2$}
\label{sec:observing_BIC_2}

The method in Sect.\,\ref{sec:observing_BIC_1} gives 
$N_{+}$ (and $N_{-}$) in a simple manner. 
However, it does not yield 
either the wave function or the eigenenergy  of the BIC.\footnote{
For example, suppose that we know  $N_{+}= 1$ and the phase shift $\delta(k)$ in 
Fig.\ref{fig:delta_T_Gauss}.   
Then, $\langle k' |V | k  \rangle$ in 
Fig.\ref{fig:inv_example_minus_1} and that in Fig.\ref{fig:inv_example_minus_4}
are equally possible as a potential of the system.
However  the former gives the eigenenergy  $K^2 = 1.0\, {\rm fm}^{-2}$
of the BIC and the latter gives $K^2 = 4.0\, {\rm fm}^{-2}$, 
showing that we cannot obtain the eigenenergy  $K^2$ from 
$\delta(k)$ and $N_{+}$ only.
}
In the following, therefore, we explain a theoretical 
access to the BIC wave function $\phi(k)$ and its eigenenergy $K^2$ 
for a given potential $\langle k' |V | k  \rangle$.\footnote{
One might expect that one could easily obtain $\phi(k)$ and $K^2$ by 
numerically diagonalizing the Hamiltonian $H=H_0 + V$ in momentum space.
This is not, however,  true;   eigenvectors one would get are   
 linear combinations of $| \phi \rangle$ and  $| K \rangle\!_+ $,
which are embedded in scattering states.
}

Let us start with a system with $N_{+}=1$ and $N_{-}=0$.
First, using  the  given potential $\langle k' |V | k  \rangle$,  
we  calculate  $\langle k' |T | k  \rangle$ by the Lippmann-Schwinger equation, 
which in turn gives  $\langle k' |V_{\rm S} | k  \rangle$ by Eq.(\ref{eqn:def_VPS}).  
Then,  with knowledge of
 $\langle k' |V_{\rm B} | k  \rangle = \langle k' |V | k  \rangle- \langle k' |V_{\rm S} | k  \rangle$,
which we know has the form of Eq.(\ref{eqn:def_VPB_single_E}),
we can easily identify the eigenenergy  $E=K^2$ and the wave function $\phi(k)$
of the BIC.

Next, let us turn to general systems with $N_{+} \geqq 1$.  
Obviously, we can  easily obtain $\langle k' |V_{\rm B} | k  \rangle$ 
as in the above.
Then, it is convenient to split $\langle k' |V_{\rm B} | k  \rangle$ of 
Eq.(\ref{eqn:def_VPB2}) as
\begin{eqnarray}
\langle k' |V_{\rm B} | k  \rangle
&=& \langle k' |V_{\rm B+} | k  \rangle + \langle k' |V_{\rm B-} | k  \rangle
\nonumber\\
&=& \sum_{i=1}^{N_{+}} (K_i^2 - {k'}^2) 
     \langle k'|\phi_i \rangle  \langle \phi_i  | k  \rangle
+ \sum_{j=N_{+}+1}^{N} (-\gamma_j^2 - {k'}^2) 
     \langle k'|\phi_j \rangle  \langle \phi_j  | k  \rangle,
\label{eqn:def_VPB_plus_minus}
\end{eqnarray}
with an obvious notation, 
where $V_{\rm B+}$ and $V_{\rm B-}$ represent 
contributions of positive and negative energy bound states, respectively.
Because we can easily obtain  $N_{-}$  negative energy bound states
  $ \{ | \phi_j \rangle,\, -\gamma_j^2  \,  ;  \, j=N_{+}+1,\cdots N\}$,    
we can calculate their contribution $\langle k' |V_{\rm B-} | k  \rangle$ 
in Eq.(\ref{eqn:def_VPB_plus_minus})
 to isolate the BIC contribution $\langle k' |V_{\rm B+} | k  \rangle$.
Then, because
$\langle k' |V_{\rm B+} | k  \rangle$ has the expression given 
by the first term in  the second line of 
Eq.(\ref{eqn:def_VPB_plus_minus}), 
it is straightforward numerically to obtain $N_{+}$ BIC's and their energies, 
$\{ \phi_i (k),\, K_i^2  \,  ;  \, i=1,\cdots N_{+}\}$.
We stress that the above analysis is feasible and clearly shows the usefulness
of the SB-decomposition  in Sect.\,\ref{sec:general_design_V}.

In the end, let us note that the above problem has been studied 
in coordinate space \cite{Mulligan_76, Mulligan_81}, to give a 
 theoretical access to  BIC's,  
$\{ \phi_i(r),\, K_i^2  \,  ;  \, i=1,\cdots N_{+}\}$.\footnote{
Here, one transforms Schr\"odinger equation  (\ref{eqn:Sch_general_coordinate})
in coordinate space into  integral equations
which incorporate boundary conditions 
for the physical and regular solutions.
Then, by looking for the zeros of the Fredholm determinants of these  integral equations,
one can derive the eigenenergies $\{ E_i=K_i^2 \}$ of the BIC's.
In the same way, by looking for the null space of the integral kernels of these equations,
one can  obtain the BIC wave functions $\{ \phi_i(r) \}$.
}
However,  we believe that our analysis in momentum space via the SB-decomposition
is much simpler and more feasible to investigate BIC's.

\section{Search for BIC}
\label{sec:quest_BIC}
Having established a general theory of BIC-supporting potentials,
we now discuss how and where we should search for a BIC in practice.
Let us recall  the generality of BIC
explained in Sect.\,\ref{sec:generality_BIC};   
if we could examine all  Hermitian potentials, most of which are nonlocal,     
we would  find as many BIC's as  negative energy bound states.
Despite great efforts,
 however, we have not yet discovered a single BIC to date.\footnote{
In condensed matter physics,
 quasi-bound states
in the continuum (QBIC's) can occur with a finite measure and therefore  
can be observed \cite{Nakamura_07}.
This is in contrast to  the BIC's   
in Sect.\,\ref{sec:discussion_local}
which are zero-measure phenomena.
In Ref.\cite{Nakamura_07}, 
one can find discussions on the relation  between the BIC and the QBIC.
}   
We believe, however,
 that the present theory provides us with a good chance to find a BIC in practice.

\subsection{How to search for BIC}
\label{sec:comparison_k_r}
Here, we outline a possible prescription 
 how to search for a BIC in practice.
Given a potential $V$, we can SB-decompose it as $V = V_{\rm S} + V_{\rm B}$
and  explain the behaviors of $V_{\rm S}$ and $V_{\rm B}$ 
both in momentum (Sect.\,\ref{sec:results_VS_plus_VB_makes_V}) and 
coordinate  (Sect.\,\ref{sec:results_VS_plus_VB_makes_V_r})  spaces.
Here, $V_{\rm B}$ has the information on 
the bound state, and therefore on the BIC, while $V_{\rm S}$ does not. 
We examine, therefore, $V_{\rm B}$ in momentum and coordinate spaces. 
For simplicity,  we consider systems with a single bound state only.

\subsubsection{Search in momentum space}
\label{sec:search_k_space}

In momentum space, the $k'$-dependence of $\langle k' |V_{\rm B} | k  \rangle$
clearly tells whether the system supports a BIC or not as explained in 
Sects.\,\ref{sec:results_VS_plus_VB_makes_V} and \ref{sec:observing_BIC_2}.\footnote{   
We can easily identify the existence of a BIC 
simply by looking into the sign of $\langle k' |V_{\rm B} | k  \rangle$ at the origin
$(k'=k=0)$.   If $\langle 0 |V_{\rm B} | 0  \rangle$ is positive, the system has a BIC.  
Further, a  node $k'=K$ of $\langle k' |V_{\rm B} | k  \rangle$ 
along the $k'$-axis gives the (positive) eigenenergy $E=K^2$ of the BIC.
}
In coordinate space, on the other hand, we cannot realize the existence of a BIC 
so clearly as in momentum space.   We have seen that 
a node of $\langle r' |V_{\rm B} | r  \rangle$ along the $r'$-axis
moves as the eigenenergy $E$ of the bound state $| \phi  \rangle$ varies.
 However, the above shift of the node does not give a clear criterion of the BIC. 

To summarize, 
we should choose the momentum space to look for a theoretical evidence of BIC
in a given potential $V$, 
i.e., we had better examine  
$\langle k' |V_{\rm B} | k  \rangle$ rather than $\langle r' |V_{\rm B} | r  \rangle$.

\subsubsection{Search in coordinate space}
\label{sec:search_r_space}

As shown in Sects.\,\ref{sec:discussion_local} and \ref{sec:discussion}, 
we have  to examine nonlocal potentials to find a BIC.
Because the nonlocality is defined in coordinate space,\footnote{
We use ``local'' and ``nonlocal'' in coordinate space only.  
} 
we can keep it under better control in coordinate space than in momentum space.\footnote{
As an example, let us consider $V_0$ in Sect.\,\ref{sec:Gaussian_potential};
its expression (\ref{eqn:gauss_r}) in coordinate space is obviously local, being
proportional to $\delta (r'-r)$, 
while Eq.(\ref{eqn:gauss_k})   in momentum space 
is not proportional to $\delta (k'-k)$.
Therefore,  it is  straightforward to distinguish ``local'' potentials from ``nonlocal'' ones
in coordinate space, while it is not the case in momentum space. 
This implies that  
the coordinate space is more suitable  than the momentum space
to control the nonlocality of a potential.
}

First,  
let us start from a system with a (generally nonlocal) potential $V_{\rm init}(r',r)$ supporting 
 a bound state $\phi(r)$ at a negative energy $-\gamma^2$.
By introducing a nonlocal perturbation $(E + \gamma^2) \phi (r') \phi(r)$,
we obtain a system with  the following potential:
\begin{equation}
 V(r',r) \, = \, V_{\rm init}(r',r) \, + \, (E + \gamma^2) \phi (r') \phi(r).
\label{eqn:transformed_nonlocal_V0}
\end{equation}
Using the  argument 
leading to Eq.(\ref{eqn:transformed_gauss_r}), 
we can show that $ V(r',r)$ of Eq.(\ref{eqn:transformed_nonlocal_V0}) now
describes $\phi(r)$ as a BIC at the positive energy $E$.
In other words, only by introducing a nonlocal perturbation   
proportional to $\phi (r') \phi(r)$, we can transform
the negative energy bound state $\phi(r)$  into a BIC.
The SB-decomposition  explains that the opposite is also true; 
any BIC-supporting potential can be obtained in this way from a potential without a BIC.\footnote{
Any BIC-supporting potential can be uniquely expressed as 
$V = V_{\rm S} + V_{\rm B}$, where $V_{\rm B}$ is given by 
$\langle k' |V_{\rm B} | k  \rangle=  (E - {k'}^2) \langle k'|\phi \rangle  \langle \phi | k  \rangle$
of Eq.(\ref{eqn:def_VPB_single_E}) with $E>0$.  By replacing the
positive $E$ by a negative one, we can relate the  potential with a BIC 
to a potential without a BIC in a unique fashion.
}

Second,  let us note the following;  the generality of BIC in Sect.\,\ref{sec:generality_BIC}  
 means that BIC's are robust against perturbations.\footnote{
This is true in the group of all Hermitian nonlocal potentials.  
If we restrict ourselves to its  subgroup of  Hermitian local potentials only,
the situation is different; 
it has been shown that the BIC of von Neumann and Wigner (local) potential, 
  is vulnerable to any local perturbation \cite{Pursey_95, Weber_98}.
}
This means that,  if a potential $V(r',r)$ supports  
 a BIC, its neighboring potential  $V(r',r)+ \delta V(r',r)$,
where $\delta V(r',r)$ is an arbitrary small deviation, 
 would also support a BIC.
In other words,  for $ V(r',r)$ of Eq.(\ref{eqn:transformed_nonlocal_V0}) 
to support a BIC, the perturbation on the right hand side only needs to behave 
 ``roughly'' as  $(E + \gamma^2) \phi (r') \phi(r)$.

We summarize the present section as follows;
if we are going to find a BIC 
in the neighborhood of a system with $V_{\rm init}(r',r)$ supporting 
a  negative energy bound state, 
it is absolutely necessary to introduce a ``nonlocal'' perturbation,  
which is roughly separable as $\phi (r') \phi(r)$, to the system.
For this purpose, we had better work in coordinate space
where ``nonlocality'' is apparent.

\subsubsection{Combined search}
\label{sec:combined_search}

Let us combine the findings in 
Sects.\,\ref{sec:search_k_space} and \ref{sec:search_r_space}
to devise  a possible scenario for the research of a BIC.

We first work in coordinate space; 
we look into every nonlocal potential $V(r',r)$ which is realizable experimentally.
Then, if we find a potential  $V(r',r)$ which can be put as
\begin{equation}
 V(r',r) \, = \, V_{0}(r',r) \, + \, \Delta V(r',r), 
\label{eqn:where_nonlocal_V}
\end{equation}
in such a way 
 (i) that $V_{0}(r',r)$ supports a bound state, 
and (ii) that   $\Delta V(r',r)$ is separable, 
we regard 
the potential $V(r',r)$ at hand as a good candidate  for a  BIC-supporting potential.

Next, we bring the above candidate $V$ into momentum space; 
we calculate its SB-decomposition $V = V_{\rm S} + V_{\rm B}$
and  look into the sign of $\langle k' |V_{\rm B} | k  \rangle$ at the origin
$(k'=k=0)$.   If we find that it is positive, we can conclude that 
the candidate $V$ at hand is truly a BIC-supporting potential.

\subsection{Where to search for BIC}
\label{sec:where}

In the end, we dare to make a conjecture,
on the basis of the result in Sect.\,\ref{sec:combined_search},
  where we should  search for a BIC.

Suppose that we have a system with $V_{0}(r',r)$ of Eq.(\ref{eqn:where_nonlocal_V})
 which supports a bound state at $E<0$.  
Then, the occurrence of a BIC depends crucially on
the introduction of a nonlocal separable perturbation  $\Delta V(r',r)$. 
This is, however,  admittedly  difficult, because  fundamental inter-particle interactions 
are described mostly by local potentials.   Nonlocal interactions usually come about  
as an ``effective'' interaction which incorporates effects
of antisymmetrization  or some degrees of freedom outside 
the model space under consideration \cite{Mulligan_76}. 
We expect, therefore, the following two scenarios might be possible.

\begin{itemize}
\item[(i)]
First, we consider the possibility to simulate the above 
 $\Delta V(r',r)$ by a change of 
the Fock potential of a many-body system.
 Here, we choose a system which is described as 
a  single particle state  $\phi(r)$ on top of a core.\footnote{
Suppose $\Nu{Ca}{41}{}$ nucleus as an example, which is
described mostly as a single neutron in $0f_{7/2}$ orbit 
 on top of $\Nu{Ca}{40}{}$ core.
}
We suppose that the core gives a mean  (generally nonlocal) potential $V_0(r',r)$
which describes $\phi(r)$ at a negative energy $-\gamma^2$.\footnote{
Here, we imagine that the Hartree-Fock mean field $V_{\rm HF}(r',r)$
plays the role of our $V_0(r',r)$ in Eq.(\ref{eqn:where_nonlocal_V}).
}
Now,  we introduce a nonlocal perturbation to $\Delta V(r',r)$ 
by changing the core.  One way might be to remove or excite a particle from the core.\footnote{
By  removing a particle in a state $\varphi(r)$ in the core, for example, 
 we change the nonlocal Fock potential  
by $\varphi (r') v(r',r) \varphi(r)$ in the standard notation.
}
If the nonlocal perturbation thus introduced has a sufficient overlap 
with $\phi (r') \phi(r)$, 
the resultant system might support a BIC.

\item[(ii)]
Second, we may use a nonlocal effective interaction 
 to generate the above  $\Delta V(r',r)$.
In nuclear physics, for example, we naturally expect that $\Lambda N$ 
effective interaction is mostly nonlocal.\footnote{
Effective interaction between  $\Lambda$ and nucleon  
 does not originate from  
a one-pion-exchange process because of the isospin conservation.
It necessarily requires at least a two-pion-exchange process 
which naturally induces  a nonlocal potential.  
}
This implies that 
we may introduce a highly nonlocal perturbation into a nucleus
by replacing a nucleon with a $\Lambda$ particle.
We expect, therefore,  
$\Lambda$-hypernuclei may be a promising field to search for a BIC.\footnote{
For $\Lambda$ and $\Lambda$-hypernuclei, see Refs.\,\cite{Hyodo_12, Gal_16}.
 }

\end{itemize}

We admit that the above scenarios are far from being well-designed.
However, we hope that they serve as a starting point to set out on a search for a BIC.


\section{summary}
\label{sec:summary}
In this work, we have presented a general theory to construct 
 a potential that supports a bound state at a positive energy, i.e., 
a bound state in the continuum (BIC) with numerical examples.
Our theory is based on the SB-decomposition of a Hermitian potential,
and is able to express any potential supporting a BIC.
The present theory has clarified  
(i) that the BIC is a rare phenomenon if we look into local potentials only, 
and astonishingly enough, 
(ii) that BIC's are as common as  negative energy bound states
if we look into all Hermitian potentials most of which are nonlocal.     
Accordingly, all BIC-supporting potentials in the present theory are nonlocal.    
In all investigations in search of a BIC, therefore, 
it is crucial to devise and examine systems with nonlocal potentials.
In the end,  
we have proposed some scenarios to search for a BIC 
in real systems with possible nonlocal potentials.
We believe that the present theory will boost research activities 
on nonlocal potentials and  the BIC in various quantum systems.







\appendix

\section{Existing theories of BIC-supporting potential}
\label{app:existing_theories}
In this appendix, we briefly explain two well-known potentials  which support a BIC,
the von Neumann and Wigner potential and the separable potential.

\subsection{von Neumann and Wigner potential}
\label{app:vonNeumann_Wigner_pot}
We outline the method of von Neumann and Wigner \cite{vonNeumann29}
following Ref.\cite{Stillinger_PRA_75}.
We consider a local potential $V(r)$ in the S-wave channel,
for which  the Schr\"odinger equation   
 in coordinate space reads
\begin{equation}
(-\nabla^2 +V(r))  \phi(r)  = E \phi(r),
\label{eqn:Sch_vNW_1}
\end{equation}
where $\nabla^2 = \dfrac{{\rm d}^2}{{\rm d}r^2} + \dfrac{2}{r}\dfrac{{\rm d}}{{\rm d}r}$.
Now we look for $V(r)$ which gives a normalizable  $\phi(r)$
at a positive energy $E$ via Eq.(\ref{eqn:Sch_vNW_1}).

Let us solve Eq.(\ref{eqn:Sch_vNW_1}) for $V(r)$:
\begin{equation}
V(r) = E + \frac{\nabla^2  \phi(r)}{\phi(r)}, 
\label{eqn:Sch_vNW_2}
\end{equation}
which explicitly shows that, to obtain a smooth $V(r)$,  each zero of  
$\phi(r)$  must be canceled by vanishing $\nabla^2  \phi(r)$.   
We note that the free state  $\phi(r) = j_{0}(kr)$ with $E= k^2$
naturally gives $V(r) = 0$
by  Eq.(\ref{eqn:Sch_vNW_2}).
Therefore,  we look for a BIC solution $\phi(r)$  
by modulating the free state $j_{0}(kr)$ as
\begin{equation}
\phi(r) = j_{0}(kr) f(r), 
\label{eqn:Sch_vNW_3}
\end{equation}
where  the envelope $f(r)$ is a smooth function without a zero.
Here, we have to require $f(r)$ be chosen in such a way (i) that
$f(r)$ guarantees $\phi(r)$ be normalizable, 
and (ii) that each zero of   $\phi(r)$  be canceled by vanishing $\nabla^2  \phi(r)$
in the right hand side of Eq.(\ref{eqn:Sch_vNW_2}).
By substituting $\phi(r) $ of Eq.(\ref{eqn:Sch_vNW_3})
into Eq.(\ref{eqn:Sch_vNW_2}), 
we obtain 
\begin{equation}
V(r) = E -k^2 + 2k \cot(kr) \frac{f'(r)}{f(r)}  + \frac{f''(r)}{f(r)}.
\label{eqn:Sch_vNW_4}
\end{equation}
Then, we see that  above condition (ii) of $\phi(r)$
has been translated into a condition of $f(r)$, i.e., 
(ii') that $\dfrac{f'(r)}{f(r)} =0$ at $kr=n\pi,\,n=0,1,\cdots$.
We can ensure condition (ii')  by choosing $f(r)$ to be a 
smooth function of
\begin{equation}
R = 4k \int_0^r {\rm d}r' \sin^2(kr') = 2kr - \sin(2kr).
\label{eqn:Sch_vNW_5}
\end{equation}
In Ref.\,\cite{vonNeumann29},  taking  condition (i) into account,  
they chose
\begin{equation}
f(r) = \frac{1}{ A^2 + R^2}, 
\label{eqn:Sch_vNW_6}
\end{equation}
where $A(\ne 0)$ is an arbitrary constant.
Note also that, for $V(r)$ of Eq.(\ref{eqn:Sch_vNW_4})
 to tend to zero as $r \to \infty$, we must have $E = k^2$. 
Summarizing the above, Eq.(\ref{eqn:Sch_vNW_4}) gives 
\begin{equation}
V(r) =  - \frac{128 A^2 k^2 \sin^4(kr)}{( A^2 + (2kr-\sin(2kr))^2)^2}  
+\frac{96  k^2 \sin^4(kr) -16k^2 (2kr-\sin(2kr)) \sin(2kr)}{ A^2 + (2kr-\sin(2kr))^2}  ,
\label{eqn:Sch_vNW_7}
\end{equation}
which behaves as
\begin{equation}
V(r) \propto r^4,~(r \to 0), ~~~ V(r) \propto \frac{\sin 2kr}{2kr},~(r \to \infty). 
\label{eqn:Sch_vNW_8}
\end{equation}
The above method of von Neumann and Wigner explicitly constructs 
 the local potential $V(r)$ of Eq.(\ref{eqn:Sch_vNW_7})
that describes $\phi(r)$ of Eq.(\ref{eqn:Sch_vNW_3}) as a BIC at $E=k^2$.

\subsection{separable potential}
\label{app:separable_pot}

It is well known that a separable potential can be tuned to support a BIC
\cite{Chadan89}.
Here, we consider the following separable potential 
 in momentum space in the S-wave channel: 
\begin{align}
&\langle k' | V  | k \rangle  = \lambda g(k') g(k), 
\label{eqn:pot_separable_1}
\end{align}
where $\lambda$ is a real coupling strength,  and $g(k)$ is a real function.

Let us assume a BIC  $\phi(k)$ at a positive energy $E=K^2$.
Then,  Schr\"odinger equation  (\ref{eqn:Sch_BIC}) takes on 
the following form:
\begin{equation}
k^2 \phi(k) +  g(k) \int \!\! \frac{p^2 {\rm d}p}{(2\pi)^3} ~ \lambda  g(p)  \phi(p)
= K^2 \phi(k),
\label{eqn:Sch_BIC_sep}
\end{equation}
which directly yields $\phi(p)$ as
\begin{equation}
\phi(p) = {\mathcal N} \frac{  g(p) } {K^2-p^2} ,
\label{eqn:pot_separable_2}
\end{equation}
where ${\mathcal N}$ is a normalization constant.
By substituting Eq.(\ref{eqn:pot_separable_2}) into 
Schr\"odinger equation  (\ref{eqn:Sch_BIC_sep}), we immediately arrive at 
\begin{equation}
\int \!\! \frac{p^2 {\rm d}p}{(2\pi)^3} ~ \frac{ \lambda g(p)^2 } {K^2-p^2} = 1.
\label{eqn:pot_separable_3}
\end{equation}
Finally,  for $\phi(p)$ of Eq.(\ref{eqn:pot_separable_2})
to be normalizable, we have to require 
\begin{equation}
g(K) = 0.
\label{eqn:pot_separable_4}
\end{equation}

The above observation shows that a BIC is a solution to the set of 
Eqs.(\ref{eqn:pot_separable_3}) and (\ref{eqn:pot_separable_4}), 
which can be found in the following manner.
First, we fix $K$ arbitrarily, and then choose a function $g(k)$ which vanishes 
 at $k=K$ so that Eq.(\ref{eqn:pot_separable_4}) is satisfied.
Second, by varying the coupling strength $\lambda$, 
we can find  $\lambda = \lambda_{\rm c}$ which satisfies
Eq.(\ref{eqn:pot_separable_3}).   The above procedure shows that, 
for  any $g(k)$ with a node at $k=K$,
we can always tune the coupling strength $\lambda$ 
 in such a way that the potential describes 
 $\phi(k)$ of Eq.(\ref{eqn:pot_separable_2}) as a BIC at $E=K^2$.
Conversely,  for a fixed $g(k)$ with a node, 
  the BIC only appears at a special value  $\lambda = \lambda_{\rm c}$
 of the coupling strength, i.e., the BIC is a rare phenomenon
for a system described by  a separable potential.


%



\let\doi\relax




\end{document}